\documentclass[12pt,onecolumn,oneside,a4paper,peerreview]{IEEEtran}

\usepackage{cite}
\usepackage{graphicx}
\usepackage{amsmath}
\usepackage{times}
\usepackage{latexsym}
\usepackage{graphicx}
\usepackage{bm}
\usepackage{algorithm}
\usepackage{algpseudocode}
\usepackage{amssymb}
\usepackage{bbm}
\usepackage[center]{caption2}
\usepackage{stfloats}
\usepackage{cases}
\usepackage{url}
\usepackage{array}
\usepackage{setspace}
\usepackage{fancyhdr}
\usepackage{color}
\usepackage{subfigure}
\usepackage{epstopdf}
\usepackage{enumerate}

\newtheorem{theorem}{Theorem}

\newtheorem{remark}{Remark}
\newtheorem{corollary}{Corollary}
\newtheorem{proposition}{Proposition}

\def\proof{\noindent\hspace{2em}{\itshape Proof: }}
\def\endproof{\hspace*{\fill}~$\square$\par\endtrivlist\unskip}

\allowdisplaybreaks[4]
\begin{document}
\title{Multiple RISs Assisted Cell-Free Networks With Two-timescale CSI: Performance Analysis and System Design}
\author{Xu Gan,~\IEEEmembership{Graduate Student Member,~IEEE}, Caijun Zhong,~\IEEEmembership{Senior Member,~IEEE}, Chongwen Huang,~\IEEEmembership{Member,~IEEE}, Zhaohui Yang,~\IEEEmembership{Member,~IEEE} and Zhaoyang Zhang,~\IEEEmembership{Member,~IEEE}
}

\maketitle

\begin{abstract} \vspace{-10pt}
Reconfigurable intelligent surface (RIS) can be employed in a cell-free system to create favorable propagation conditions from base stations (BSs) to users via configurable elements. However, prior works on RIS-aided cell-free system designs mainly rely on the instantaneous channel state information (CSI), which may incur substantial overhead due to extremely high dimensions of estimated channels. To mitigate this issue, a low-complexity algorithm via the two-timescale transmission protocol is proposed in this paper, where the joint beamforming at BSs and RISs is facilitated via alternating optimization framework to maximize the average weighted sum-rate. Specifically, the passive beamformers at RISs are optimized through the statistical CSI, and the transmit beamformers at BSs are based on the instantaneous CSI of effective channels. In this manner, a closed-form expression for the achievable weighted sum-rate is derived, which enables the evaluation of the impact of key parameters on system performance. To gain more insights, a special case without line-of-sight (LoS) components is further investigated, where a power gain on the order of $\mathcal{O}(M)$ is achieved, with $M$ being the BS antennas number. Numerical results validate the tightness of our derived analytical expression and show the fast convergence of the proposed algorithm. Findings illustrate that the performance of the proposed algorithm with two-timescale CSI is comparable to that with instantaneous CSI in low or moderate SNR regime. The impact of key system parameters such as the number of RIS elements, CSI settings and Rician factor is also evaluated.  Moreover, the remarkable advantages from the adoption of the cell-free paradigm and the deployment of RISs are demonstrated intuitively.

\begin{center}
{\bf Index Terms}
\end{center}
Cell-free system, reconfigurable intelligent surface (RIS), two-timescale beamforming, statistical CSI, low-complexity algorithm.

\end{abstract}

\newpage
\section{Introduction}
The proliferation of mobile phones and other portable devices continuously exacerbates the demand for data transmission in wireless networks. To support the mounting data traffic growth, the cell-free system was introduced in \cite{cellfree1,cellfree2,cellfree3}, which has attracted extensive research interest due to its high spectral efficiency\cite{cellfree6}. Nevertheless, the traditional cell-free paradigm requires a large-scale deployment of BSs to guarantee favorable performance, leading to an unsatisfying energy efficiency performance due to enormous hardware and power expense \cite{cellfree7}.

Fortunately, an emerging technology named reconfigurable intelligent surface (RIS) \cite{RIS1,RIS2} has been identified as a solution to address the above problems by creating favorable propagation conditions from BSs to UEs with low-cost and power-efficient elements. As such, the RIS has been investigated in various aspects and under different setups \cite{joint1,joint2,learning1,learning2,learning3,RIS3,RIS4}. For instance, the joint optimization of beamformers at the BS and RIS has been widely studied in \cite{joint1,joint2}. Capitalizing on the recent advancements in artificial intelligence (AI), the deep reinforcement learning is used to tackle the RIS phase-shift design \cite{learning1}, channel estimation \cite{learning2}, secure communication \cite{learning3}, etc. {Moreover, since RIS does not require extra hardware implementation \cite{RIS6}, it is natural to envision a cell-free system integrating RISs, which can reap both advantages of these two technologies. Hence, compared with the conventional cell-free system, a lower level of power consumption is required to achieve a satisfying performance. In other words, the fusion of RISs into a cell-free system increases degrees of freedom to enhance the performance with low cost and power consumption. In this regard, there have been several preliminary explorations on RIS-aided cell-free systems \cite{cellfree-ris1,cellfree-ris2,cellfree-ris3}. Specifically, authors in \cite{cellfree-ris1} formulated the joint beamforming design at the BSs and RISs in the wideband scenario to maximize the network capacity, while the work in \cite{cellfree-ris2} aimed to maximize the energy efficiency. In addition, \cite{cellfree-ris3} proposed a fully decentralized design framework for cooperative beamforming.}

From the aforementioned works in RIS-aided cell-free systems, the instantaneous channel state information (CSI) of all links is assumed to be perfectly known at BSs. However, this is impractical, since accurate CSI acquisition may incur significant overhead due to a large number of RIS elements \cite{estimation1}. Responding to this, several works have proposed the idea of using statistical CSI\footnote{{The statistical CSI refers to the information for LoS components, the path-loss coefficients and Rician $K$ factors, which can be viewed as the constant in several coherence intervals.}} for the design of RIS-assisted systems \cite{S-CSI1,S-CSI2,S-CSI3}. However, the utilization of statistical CSI for both active and passive beamformers may severely degrade the system performance. As such, a novel countermeasure named two-timescale transmission protocol \cite{tts} is quite suitable for the RIS-aided cell-free systems. On the one hand, the passive beamformers at RISs are optimized by exploiting the statistical CSI, without having to acquire CSI in every time slot. On the other hand, the active beamformers at BSs are designed through the instantaneous CSI of effective channels, where the process of channel estimation is almost the same as that in conventional multiple-input-multiple-output (MIMO) systems. While the potential of the two-timescale beamforming in the RIS-aided cell-free system is conceivable, fundamental understanding and design guidelines are still lacking. Motivated by this, we propose an effective and low-complexity solution to design this system. 

The main contributions of this paper can be summarized as follows:
\begin{itemize}
\item The closed-form expression of the achievable weighted sum-rate is derived, which facilitates the understanding on the impact of key system parameters, such as Rician $K$ factor, the number of BS antennas, and the number of RIS elements on the achievable rate. Furthermore, we investigate a special case without line-of-sight (LoS) components to gain more insights, in which the weighted sum-rate increases logarithmically with $\mathcal{O}(M)$, with $M$ being the number of BS antennas. Therefore, the system will have significant benefits offered by the adoption of a large number of BS antennas.	
	
\item An achievable weighted sum-rate maximization problem under the two-timescale transmission protocol is formulated and decomposed into two subproblems via an alternating optimization framework. Specifically, a penalty dual decomposition (PDD)-based method is conceived to optimize RISs beamformers based on the statistical CSI, while a primal dual subgradient (PDS)-based method is proposed to design BSs beamformers. Moreover, theoretical analyses on the properties of the proposed algorithm, i.e., the convergence behavior and the computational complexity, are also derived.

\item Finally, simulation results are presented to validate the tightness of our derived analytical expression and show the fast convergence of our proposed algorithm. Findings illustrate that the performance of the proposed algorithm with two-timescale CSI is comparable to that with instantaneous CSI in low or moderate SNR regimes. The impact of key system parameters such as the number of RIS elements, CSI settings and Rician factor is also evaluated. Moreover, the remarkable advantages from the adoption of the cell-free paradigm and the deployment of RISs are demonstrated intuitively.
\end{itemize}

\subsection{Structure and Notations}
The rest of this paper is organized as follows. In Section~\ref{model}, we present the channel model and transmission protocol. The closed-form expression of the achievable weighted sum-rate and a special case without LoS components are analyzed in Section~\ref{analysis}. Then, the average weighted sum-rate maximization problem is formulated and the design of RISs phase shifts and BSs power allocation coefficients are elaborated in Section~\ref{design}. In Section~\ref{simulation}, numerical results are provided to validate the tightness of analytical expressions, evaluate the proposed algorithm performance and demonstrate the impacts of key parameters. Finally, we conclude the paper in Section~\ref{conclusion}.

{\it Notation}: We use bold lower case letters to denote vectors and lower case letters to denote scalars. $\text{Re}\{ \cdot \}$ represents the real part of a complex value. $\mathbf{I}_N$ denotes an identity matrix with subscript $N$ being the matrix dimension. The operators $\mathbb{E}\{ \cdot \}$, $\text{tr}(\cdot)$ and $\|\cdot\|$ stand for the expectation, trace and Euclidean norm operations, respectively. The superscripts $(\cdot)^{*}$, $(\cdot)^{T}$ and $(\cdot)^H$ are denoted as the conjugate, transpose, and conjugate-transpose operations, respectively. $\mathcal{CN}(\mathbf{a},\mathbf{B})$ represents the symmetric complex-valued Gaussian distribution with mean $\mathbf{a}$ and convariance matrix $\mathbf{B}$. The operation $\text{diag}(\mathbf{x})$ generates a diagonal matrix with the elements of $\mathbf{x}$ along its main diagonal, while $\text{diag}(\mathbf{X})$ means the block diagonal operation. In addition, the operation $\text{Diag}(\mathbf{A})$ returns a column vector of the main diagonal elements of $\mathbf{A}$.

\section{System Model}\label{model}
\begin{figure}[ht]
	\centerline{\includegraphics[width=0.47\textwidth]{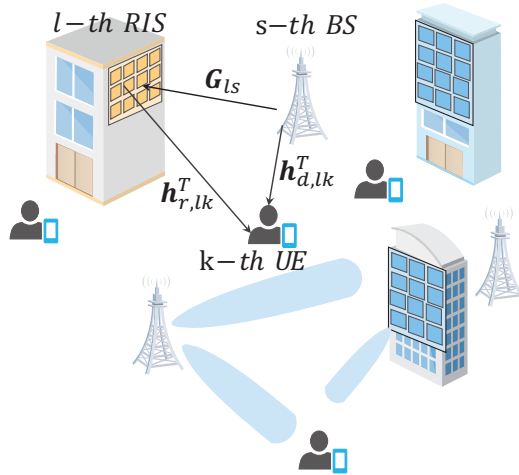}}
	\caption{The illustration of the RIS-aided cell-free channels.}\label{system}
\end{figure}

We consider a RIS-aided cell-free system as illustrated in Fig. \ref{system}, where $K$ single-antenna UEs are served by $S$ BSs with the aid of $L$ RISs. Furthermore, the $s$-th BS ($s \in \{1,...,S\}$) is equipped with $M$ antennas arranged in the form of uniform linear array (ULA) and the $l$-th RIS ($l \in \{1,...,L\}$) is equipped with $N$ ($N=N_r \times N_c$) reflecting elements arranged in the form of uniform planar array (UPA) with length $N_r$ and width $N_c$. The channels from the $s$-th BS to the $l$-th RIS, from the $s$-th BS to the $k$-th UE $(k \in \{ 1,...,K\})$, and from the $l$-th RIS to the $k$-th UE are denoted by $\mathbf{G}_{ls} \in \mathbb{C}^{N \times M}$, $\mathbf{h}_{d,ks}^T \in \mathbb{C}^{1 \times M}$ and $\mathbf{h}_{r,lk}^T \in \mathbb{C}^{1 \times N}$, respectively. Then, the effective channel from the $s$-th BS to the $k$-th UE can be expressed by
\vspace{-4mm}
\begin{equation}\vspace{-3mm}
\mathbf{h}_{ks}^T = \sum_{l=1}^L \mathbf{h}_{r,lk}^T \bm{\Theta}_l \mathbf{G}_{ls} + \mathbf{h}_{d,ks}^T
\end{equation}
where $\bm{\Theta}_l$ is the phase-shift matrix of the $l$-th RIS and $\bm{\Theta}_l = \text{diag}(e^{j\phi_{l,1}},...,e^{j\phi_{l,N}})$ with $\phi_{l,n}$ being the phase of the $n$-th element in the $l$-th RIS. A central precoding unit (CPU) is deployed to connect all BSs and RISs through backhaul links \cite{backhaul1}.
\vspace{-3mm}

For downlink transmission, the $s$-th BS broadcasts the superimposed signal $\mathbf{x}=\sum_{k=1}^K \mathbf{w}_{ks} u_{k}$, and the received signal at the $k$-th UE can be expressed as
\begin{equation}\vspace{-1mm}
	y_k = \sum_{s=1}^S \mathbf{h}_{ks}^T \sum_{k=1}^K \mathbf{w}_{ks} u_{k} + \mathbf{n}_k,
\end{equation}
where $u_{k}$ is the information bearing symbol for the $k$-th UE; $\mathbf{w}_{ks}$ denotes the precoding vector for the $s$-th BS to the $k$-th UE, subject to the transmit power constraint $\sum_{k=1}^K \| \mathbf{w}_{k,s} \|^2 \leq P_{s,\max}$ with $P_{s,\max}$ being the maximum transmit power at the $s$-th BS; $\mathbf{n}_k$ is the additive white Gaussian noise and $N_0$ is the noise density power.

Then, the achievable rate of the $k$-th UE is given by
\begin{equation}
R_k = \log_2 \left( 1+\frac{\| \sum_{s=1}^S \mathbf{h}_{ks}^T \mathbf{w}_{ks}\|^2}{\sum_{i \neq k}^K \|\sum_{s=1}^S \mathbf{h}_{ks}^T \mathbf{w}_{is}\|^2+N_0} \right).
\end{equation}

\subsection{Channel Model}
We assume that RISs are deployed at positions with line-of-sight and non-line-of-sight (NLoS) paths to both BSs and UEs as in \cite{S-CSI1,tang_model}. Hence, the Rician distribution is used to model these channels, i.e.,
\begin{equation}\vspace{-1mm}
\mathbf{G}_{ls} = \sqrt{\beta_{ls}\frac{K_{ls}}{K_{ls}+1}}\mathbf{\bar{G}}_{ls}+\sqrt{\beta_{ls} \frac{1}{K_{ls}+1}}\mathbf{\tilde{G}}_{ls}
\end{equation}
\begin{equation}\vspace{-1mm}
\mathbf{h}_{r,lk} = \sqrt{\beta_{r,lk}\frac{K_{r,lk}}{K_{r,lk}+1}}\mathbf{\bar{h}}_{r,lk}+\sqrt{\beta_{r,lk} \frac{1}{K_{r,lk}+1}}\mathbf{\tilde{h}}_{r,lk}
\end{equation}
and
\begin{equation}\vspace{-2mm}
\mathbf{h}_{d,ks}=\sqrt{\beta_{d,ks}\frac{K_{d,ks}}{K_{d,ks}+1}}\mathbf{\bar{h}}_{d,ks}+\sqrt{\beta_{d,ks}\frac{1}{K_{d,ks}+1}}\mathbf{\tilde{h}}_{d,ks},
\end{equation}
where $K_{ls}$, $K_{r,lk}$ and $K_{d,ks}$ are the Rician $K$ factors; $\beta_{ls}$, $\beta_{r,lk}$ and $\beta_{d,ks}$ are the large scale path-loss coefficients. Besides, $\mathbf{\bar{G}}_{ls}$, $\mathbf{\bar{h}}_{r,lk} $ and $\mathbf{\bar{h}}_{d,ks}$ are LoS components of the corresponding channels, which can be expressed by the array response of the ULA and UPA. Namely, we have $\mathbf{\bar{G}}_{ls} = \mathbf{a}_{N}^T(\theta_{AoA1,ls}, \phi_{AoA1,ls})\mathbf{a}_{M}(\theta_{AoD,ls})$, $ \mathbf{\bar{h}}_{r,lk} = \mathbf{a}_{N}^T(\theta_{AoA2,lk}, \phi_{AoA2,lk})$ and $\mathbf{\bar{h}}_{d,ks} = \mathbf{a}_{M}^T(\theta_{AoA3,ks})$, where $\theta_{AoA1,l}$ and $\theta_{AoA2,l}$ are the azimuth angles of arrival (AoA) to the $l$-th RIS, $\phi_{AoA1,l}$ and $\phi_{AoA2,l}$ are the elevation AoAs to the $l$-th RIS, $\theta_{AoA3,s}$ is the azimuth AoA to the $s$-th BS, and $\theta_{AoD,s}$ is the azimuth angle of departure (AoD) from the $s$-th BS. Besides, $\mathbf{a}_{M}(\theta) = [1,e^{j2\pi\frac{d_1}{\lambda}\sin(\theta)},...,e^{j2\pi(M-1)\frac{d_1}{\lambda}\sin(\theta)}]$ and $\mathbf{a}_{N}(\theta,\phi) = [1,...,e^{j2\pi\frac{d_2}{\lambda}(n_r\cos(\theta)\sin(\phi)+n_c\sin(\theta)\sin(\phi))},...,\\e^{j2\pi\frac{d_2}{\lambda}((N_r-1)\cos(\theta)\sin(\phi)+(N_c-1)\sin(\theta)\sin(\phi))}]$ are the the array response with $d_1$ being the BS antenna spacing and $d_2$ being the RIS elements spacing, where $\lambda$ is the wavelength. On the other hand, $\mathbf{\tilde{G}}_{ls}$,  $\mathbf{\tilde{h}}_{r,lk}$, and $\mathbf{\tilde{h}}_{d,ks}$ are the NLoS components, whose elements are independently and identically distributed (i.i.d.) complex Gaussian random variables with zero mean and unit variance.

\vspace{-5mm}
\subsection{Transmission Protocol}
{Prior investigations on RIS-aided cell-free systems are mainly based on the pure instantaneous CSI transmission protocol \cite{cellfree-ris1, cellfree-ris2, cellfree-ris3}. However, due to a large number of RIS elements, the acquisition of the instantaneous CSI of RIS-related links will incur heavy training costs and overhead. Besides, the burdens of the channel estimation process will inevitably introduce the delay every coherent interval.}

{To avoid the huge overhead and delay caused by the channel estimation process, \cite{S-CSI1} has studied the pure statistical CSI transmission protocol in the RIS-aided MISO system. Then, the processing of channel estimation and system design is performed only once at the beginning of a time frame. This will indeed greatly reduce the complexity of system optimization and the delay, but it will also bring performance degradation significantly.}

{In order to solve the shortcomings of the above two transmission mechanisms, the two-timescale transmission protocol \cite{tts} is introduced to reduce the channel estimation overhead effectively without sacrificing much performance. On the one hand, since the number of RIS elements is much larger than that of transmit antennas at the BS, the effective CSI usually has a much smaller dimension than the RIS-related CSI. Therefore, the channel estimation overhead can be significantly reduced by exploiting the statistical CSI and the effective instantaneous CSI. On the other hand, we use the effective instantaneous CSI for active beamformers at BSs with reaping substantial performance gain.}

\begin{figure}[ht]\vspace{-2mm}
	\centerline{\includegraphics[width=0.6\textwidth]{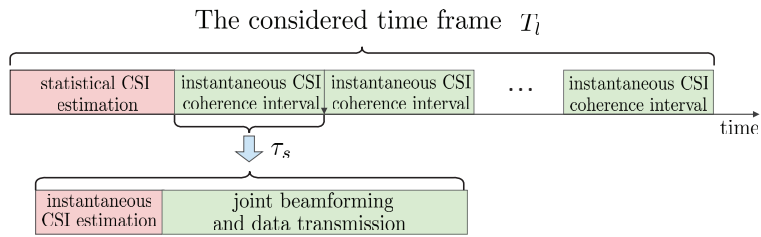}}
	\caption{The two-timescale frame structure.}\label{tts}
\end{figure}

{As illustrated in Fig. (\ref{tts}), we focus on one time frame within which the statistical CSI remains unchanged. Let $T_l$ denote the considered time frame, which can be divided into two transmission phases. Specifically, in the \textit{first phase}, the statistical CSI of all links is estimated using pilots by standard mean and convariance matrices estimation methods \cite{scsi_estimation} and fed back to BSs. Then, the \textit{second phase} consists of several instantaneous CSI coherence intervals, $\tau_s$. In each interval, the effective instantaneous CSI from UEs to BSs is first obtained by traditional channel estimation methods \cite{estimation_MIMO1, estimation_MIMO2}, and then the joint optimization framework is developed to design the beamformers at BSs and RISs with the two-timescale CSI.}

\vspace{-8mm}
\section{Achievable Sum-rate Analysis}\label{analysis}
In this section, we first present an analytical investigation of the achievable sum-rate of the system. Without loss of generality, we focus on the achievable sum-rate of the $k$-th UE. To avoid CSI exchange between distributed BSs and simplify the design, MR precoding \cite{MR_local} is adopted at BSs. Then, the precoding vector can be written as $\mathbf{w}_{ks}=\sqrt{\eta_{ks}} \mathbf{h}_{ks}^*$, satisfying the power constraint $\sum_{k=1}^K \eta_{ks} {\|\mathbf{h}_{ks}\|^2}\leq P_{s,\max}$. Thus, the achievable sum-rate of the $k$-th user can be written as\vspace{-4mm}
\begin{equation}\label{C_k}\vspace{-2mm}
	C_k = \mathbb{E} \bigg{\{}  \log_2 \big{(} 1+\frac{\| \sum\limits_{s=1}^S \sqrt{\eta_{ks}} \|\mathbf{h}_{ks}\|^2 \| ^2}{\sum\limits_{i \neq k}^K \|\sum\limits_{s=1}^S \sqrt{\eta_{is}} \mathbf{h}_{ks}^T \mathbf{h}_{is}^{*} \|^2+N_0} \big{)} \bigg{\}},
\end{equation}
and we have the following important result.

\begin{theorem} \label{theorem1}
	The ergodic capacity of $k$-th user can be approximated by
	\begin{equation}\label{capacity}
		\bar{C}_k =  \log_2 \bigg{(} 1+\frac{ \mathcal{A}_k }{ \sum\limits_{i \neq k}^K \mathcal{B}_{k,i} +N_0} \bigg{)}
	\end{equation}\vspace{-2mm}
	where
	\begin{align}
	&\mathcal{A}_k= \sum\limits_{s=1}^{S} \sum\limits_{t=1}^S \sqrt{\eta_{ks} \eta_{kt}} \big{[}  \|\bar{\mathbf{h}}_{ks}\|^2  \|\bar{\mathbf{h}}_{kt}\|^2 +2 \|\bar{\mathbf{h}}_{ks}\|^2 M \chi_{kt} +M^2 \chi_{ks} \chi_{kt}+ 2 \text{Re}\sum\limits_{l=1}^L ( a_{ls} b_{r,lk}^2 a_{lt} \bar{\mathbf{h}}_{ks}^H \bar{\mathbf{G}}_{ls}^T \bar{\mathbf{G}}_{lt}^* \notag\\
	& \bar{\mathbf{h}}_{kt}+2M a_{ls} b_{r,lk}^2 b_{lt}^2 a_{r,lk} \bar{\mathbf{h}}_{ks}^H \bar{\mathbf{G}}_{ls}^T \bm{\Theta}_l^T \bar{\mathbf{h}}_{r,lk} +  M^2 N  b_{ls}^2 a_{r,lk}^2 b_{lt}^2 b_{r,lk}^2) + N M^2 \textstyle\sum\limits_{l=1}^L  b_{ls}^2b_{r,lk}^4b_{lt}^2+ \textstyle\sum\limits_{l=1}^L \sum\limits_{m=1}^L a_{ls} a_{ms}\notag\\
	& a_{mt} a_{lt} b_{r,lk}^2 b_{r,mk}^2  \text{tr}(\bar{\mathbf{G}}_{ls}^* \bar{\mathbf{G}}_{ms}^T \bar{\mathbf{G}}_{mt}^* \bar{\mathbf{G}}_{lt}^T) \big{]} +\sum\limits_{s=1}^S \eta_{ks}\big{[}M \chi_{ks}^2 + \|\bar{\mathbf{h}}_{ks}\|^2 (b_{d,ks}^2 +N \alpha_{1,ks}+N \alpha_{3,ks})-N^2M\notag\\
	&\alpha_{2,ks}^2+N^2M\alpha_{1,ks}\alpha_{3,ks} + MN \textstyle\sum\limits_{l=1}^L b_{ls}^4 b_{r,lk}^4\big{]},
	\end{align}
	\begin{align}
	&\mathcal{B}_{k,i}  =  \sum\limits_{s=1}^{S} \sum\limits_{t=1}^S  \sqrt{\eta_{is} \eta_{it}}  \big{[} \bar{\mathbf{h}}_{ks}^T \bar{\mathbf{h}}_{is}^* \bar{\mathbf{h}}_{it}^T \bar{\mathbf{h}}_{kt}^* +\sum\limits_{l=1}^L a_{ls} a_{lt} (b_{r,li}^2 \bar{\mathbf{h}}_{ks}^T \bar{\mathbf{G}}_{ls}^H \bar{\mathbf{G}}_{lt} \bar{\mathbf{h}}_{kt}^*+b_{r,lk}^2 \bar{\mathbf{h}}_{it}^T\bar{\mathbf{G}}_{lt}^H \bar{\mathbf{G}}_{ls} \bar{\mathbf{h}}_{is}^*)+2M \notag\\
	&\text{Re}  \textstyle\sum\limits_{l=1}^L b_{lt}^2(a_{r,li} a_{r,lk}\bar{\mathbf{h}}_{ks}^T \bar{\mathbf{h}}_{is}^* \bar{\mathbf{h}}_{r,li}^T \bar{\mathbf{h}}_{r,lk}^*+ b_{r,li}^2 a_{r,lk} a_{ls} \bar{\mathbf{h}}_{ks}^T \bar{\mathbf{G}}_{ls}^H \bm{\Theta}_l^H \bar{\mathbf{h}}_{r,lk}^* +  b_{r,lk}^2 a_{r,li} a_{ls} \bar{\mathbf{h}}_{is}^T \bar{\mathbf{G}}_{ls}^H \bm{\Theta}_l^H \bar{\mathbf{h}}_{r,li}^* ) +\notag\\
	&M^2 N  \textstyle\sum\limits_{l=1}^L b_{ls}^2 b_{lt}^2 ( b_{r,lk}^2  b_{r,li}^2 +b_{r,lk}^2 a_{r,li}^2 + a_{r,lk}^2 b_{r,li}^2 ) + 2N M^2\cdot \sum\limits_{l=1}^L  b_{ls}^2 b_{r,lk}^2 b_{r,li}^2 a_{lt}^2 + \sum\limits_{l=1}^L \sum\limits_{m=1}^La_{ls} a_{lt} a_{ms} a_{mt}\notag\\
	& b_{r,lk}^2 b_{r,mi}^2  \text{tr}(\bar{\mathbf{G}}_{ls} \bar{\mathbf{G}}_{ms}^H \bar{\mathbf{G}}_{mt} \bar{\mathbf{G}}_{lt}^H)+M^2 a_{r,lk} a_{r,li} a_{r,mk} a_{r,mi} b_{ls}^2  b_{mt}^2 \bar{\mathbf{h}}_{r,lk}^T \bar{\mathbf{h}}_{r,li}^* \bar{\mathbf{h}}_{r,mi}^T \bar{\mathbf{h}}_{r,mk}^*\big{]} + \textstyle\sum\limits_{s=1}^S \eta_{is} \big{[} \|\bar{\mathbf{h}}_{ks}\|^2 \notag\\
	&(b_{d,is}^2+N \alpha_{1,is}+N \alpha_{3,is}) + \|\bar{\mathbf{h}}_{is}\|^2 (b_{d,ks}^2+N \alpha_{1,ks}+N \alpha_{3,ks}) + M \chi_{ks} \chi_{is} -N^2 M \alpha_{2,ks} \alpha_{2,is}  \big{]},
	\end{align}
	 $a_{ls}=\sqrt{\frac{\beta_{ls} K_{ls}}{K_{ls}+1}}$, $b_{ls}=\sqrt{\frac{\beta_{ls}}{K_{ls}+1}}$, $a_{r,lk}=\sqrt{\frac{\beta_{r,lk} K_{r,lk}}{K_{r,lk}+1}}$, $b_{r,lk}=\sqrt{\frac{\beta_{r,lk}}{K_{r,lk}+1}}$, $a_{d,ks}=\sqrt{\frac{\beta_{d,ks} K_{d,ks}}{K_{d,ks}+1}}$, $b_{d,ks}=\sqrt{\frac{\beta_{d,ks}}{K_{d,ks}+1}}$, $\alpha_{1,ks}=\sum_{l=1}^L b_{ls}^2 a_{r,lk}^2$, $\alpha_{2,ks}=\sum_{l=1}^L a_{ls}^2 b_{r,lk}^2$, $\alpha_{3,ks}=\sum_{l=1}^L b_{ls}^2 b_{r,lk}^2$, $\chi_{ks}=b_{d,ks}^2+N(\alpha_{1,ks}+\alpha_{2,ks}+\alpha_{3,ks})$ and $\bar{\mathbf{h}}_{ks}=\sum_{l=1}^L a_{ls} a_{r,lk} \bar{\mathbf{G}}_{ls}^T \bm{\Theta}_l^T \bar{\mathbf{h}}_{r,lk}+a_{d,ks}\bar{\mathbf{h}}_{d,ks}$.
\end{theorem}
\proof
See Appendix \ref{ergodic_capacity}.
\endproof

{Theorem 1 presents a closed-form expression of the achievable sum-rate, which quantifies the impact of key system parameters, such as the antenna number of BSs, RISs and UEs, and Rician $K$ factors. However, due to the involved expression, it is difficult to extract physical insights. Hence, we now investigate a special case without LoS components, and get the following corollary.}
\begin{corollary}
	In the case without LoS paths, the achievable sum-rate in Eq.~(\ref{capacity}) reduces to
	\begin{equation}\vspace{-2mm}
		\bar{C}_k^{NLoS} =  \log_2 \bigg{(} 1+\frac{ \mathcal{A}_k^{NLoS} }{ \mathcal{B}_{k,i}^{NLoS}+N_0} \bigg{)}
	\end{equation}
	where $\mathcal{A}_k^{NLoS}= \sum\limits_{s=1}^{S} \sum\limits_{t=1}^S \sqrt{\eta_{ks} \eta_{kt}} \big{[}N M^2 \sum\limits_{l=1}^L b_{ls}^2 b_{r,lk}^4 b_{lt}^2 +  M^2 (b_{d,ks}^2+N\alpha_{3,ks})(b_{d,kt}^2 +N\alpha_{3,kt})\big{]} +\sum\limits_{s=1}^S \eta_{ks}\big{[}M\\ (b_{d,ks}^2+N \alpha_{3,ks})^2+ MN \sum\limits_{l=1}^L b_{ls}^4 b_{r,lk}^4 \big{]}$ and $\mathcal{B}_{k,i}^{NLoS} =  M^2 N\sum\limits_{s=1}^{S} \sum\limits_{t=1}^S \sqrt{\eta_{is} \eta_{it}} \sum\limits_{l=1}^L b_{ls}^2 b_{lt}^2  b_{r,lk}^2  b_{r,li}^2 + M \sum\limits_{s=1}^S \eta_{is} (b_{d,ks}^2 \\ +N\alpha_{3,ks})(b_{d,is}^2+N\alpha_{3,is})$.
\end{corollary}
\proof
The result can be obtained by setting $K_{ls}=K_{r,lk}=K_{d,ks}=0$.
\endproof

It is observed that the achievable sum-rate is logarithmically increasing with $M$ at high SNR, which implies the great benefit of using a large number of BS antennas. Besides, as the number of RIS elements becomes larger, the desired signal power and interference power are both on the order of $\mathcal{O}{(N^2)}$, which means that increasing the RIS elements can not always improve the performance. Moreover, the achievable sum-rate becomes independent of RIS phase shift matrices, which demonstrates that it is not necessary to design the RIS beamformers in this case.
\vspace{-4mm}
\section{Joint Active and Passive Beamfomers Design}\label{design}\vspace{-2mm}

\subsection{Problem Formulation}
We consider the problem of joint beamformers design at BSs and RISs to maximize the average weighted sum-rate. Therefore, the optimization problem can be formulated as

\begin{subequations}\label{OF1}\vspace{-7mm}
	\begin{equation}
	\begin{aligned}
	(P1) \quad \max_{\mathbf{\Phi}} \mathbb{E} \bigg{\{} \max_{\bar{\bm{\eta}}} \sum\limits_{k=1}^K \mu_k \log_2 \bigg{(} 1+\frac{ \| \sum\limits_{s=1}^S \sqrt{\eta_{ks}} \|\mathbf{h}_{ks}\|^2 \| ^2 }{  \sum\limits_{i \neq k}^K \|\sum\limits_{s=1}^S \sqrt{\eta_{is}} \mathbf{h}_{ks}^T \mathbf{h}_{is}^{*} \|^2 +N_0} \bigg{)} \bigg{\}}
	\end{aligned}
	\end{equation}\vspace{-6mm}
	\begin{equation}\vspace{-2mm}
	 s.t. \quad  \sum_{k=1}^K \eta_{ks} \|\mathbf{h}_{ks}\|^2 \leq P_{s,\max}, \quad s = \{1,...,S\}, \qquad  \qquad
	\end{equation}
	\begin{equation}\vspace{-4mm}
	|\bm{\Theta}_l(i,i)|=1, \quad l = \{1,...,L\},\ i=\{1,...,N\}  \quad
	\end{equation}
\end{subequations}
where $\bm{\Phi}=\text{diag}\{\bm{\Theta}_1,...,\bm{\Theta}_L\}$, $\bar{\bm{\eta}} = [\sqrt{\eta_{11}},...,\sqrt{\eta_{KS}}]$ and $\mu_k$ represents the priority weight of the $k$-th UE. In the next subsections, we will elaborate the proposed algorithm with two-timescale CSI via the alternative optimization framework.
\vspace{-4mm}
\subsection{Phase Shifts at the RISs}\vspace{-1mm}
The optimization of the RISs beamformers can be formulated based on the closed-form expression of the achievable sum-rate derived in \emph{Theorem 1}, which can be written as
\begin{equation}
\begin{aligned}
(P2) \quad &\max_{\mathbf{\Phi}} \ \sum\limits_{k=1}^K \mu_k \log_2 \bigg{(} 1+\frac{ \mathcal{A}_k }{ \sum\limits_{i \neq k}^K \mathcal{B}_{k,i} +N_0} \bigg{)}\\
& s.t. \quad |\bm{\Theta}_l(i,i)|=1, \quad l = \{1,...,L\}, \ i = \{1,...,N\}.
\end{aligned}
\end{equation}

Due to the non-convexity of $(P2)$, we first exploit the relationship between the mean-squared error (MSE) and the signal to interference plus noise ratio (SINR) established in \cite{MMSE1,MMSE2}. Specifically, assuming that the minimum mean-squared error (MMSE) receiver $r_k$ is adopted, let $\mathbb{E}\{\text{error}_k\}$ denote the $k$-th average MSE, given by\vspace{-3mm}
\begin{equation}\label{MSE}\vspace{-2mm}
	\begin{aligned}
		&\mathbb{E}\{\text{error}_k\} = \mathbb{E}_{\mathbf{u}, \mathbf{n},\mathbf{h}} \{ | r_k y_k - u_k |^2 \} \\
		&= |r_k|^2(\mathcal{A}_k+\sum_{i \neq k}^K \mathcal{B}_{k,i})-2\text{Re}\{\sum\limits_{s=1}^S \sqrt{\eta_{ks}}(\|\bar{\mathbf{h}}_{ks}\|^2+ M \chi_{ks})\}+|r_k|^2N_0+1,
	\end{aligned}
\end{equation}\vspace{-2mm}
where
\begin{equation}\label{r_k}
r_k = \frac{\sum\limits_{s=1}^S \sqrt{\eta_{ks}}(\|\bar{\mathbf{h}}_{ks}\|^2+ M \chi_{ks})}{\mathcal{A}_k+\sum\limits_{i \neq k}^K \mathcal{B}_{k,i} + N_0}.
\end{equation}\vspace{-8mm}

Then, the following result establishes the equivalence between the achievable weighted sum-rate maximization problem and the average weighted MSE minimization problem.
\begin{proposition}
The problem $(P2)$ can be reformulated equivalently as\vspace{-3mm}
\begin{equation}
	\begin{aligned}
		(P3) \quad&  \min_{\mathbf{\Phi}} \quad \sum_{k=1}^K
		\kappa_k \mu_k \mathbb{E}\{\text{error}_k\} \\
		&|\bm{\Theta}_l(i,i)|=1, \quad l = \{1,...,L\},\ i = \{1,...,N\},
	\end{aligned}\vspace{-4mm}
\end{equation}
where $\kappa_k = \big{[}|r_k|^2(\mathcal{A}_k+\sum_{i \neq k}^K \mathcal{B}_{k,i})-2\text{Re}\{\sum\limits_{s=1}^S \sqrt{\eta_{ks}}(\|\bar{\mathbf{h}}_{ks}\|^2+ M \chi_{ks})\}+|r_k|^2N_0+1\big{]}^{-1}$.
\end{proposition}
\proof
See Appendix \ref{MMSE_eq}.
\endproof

{At this stage, there are two main challenges in solving problem $(P3)$. \emph{Firstly}, the objective function is quite complex and implicit w.r.t. $\bm{\Phi}$. \emph{Secondly}, unit-modulus constraints on RIS phase-shift matrices further make the problem intractable. As a matter of fact, $(P3)$ is clarified as the class of NP-hard problem.}

{To circumvent these challenges, we employ the PDD-based method proposed in \cite{PDD} to obtain an effecient solution. Let the vector $\mathbf{u}$ denote all the RIS phase shifts, where $\mathbf{u}=[\mathbf{v}_1^T, \mathbf{v}_2^T,...,\mathbf{v}_L^T]$ and $\mathbf{v}_l=\text{Diag}(\bm{\Theta}_l)$. To further proceed, we have the following reformulation.}
\begin{proposition}
The problem $(P3)$ can be reformulated equivalently as\vspace{-2mm}
\begin{equation}\vspace{-6mm}
	\begin{aligned}
		(P4) \quad & \min_{\mathbf{u}, \mathbf{x}, \mathbf{z}} \quad \mathcal{F}(\mathbf{u}, \mathbf{x})\\
		& s.t. \quad \mathbf{x}=\mathbf{u}, \ \mathbf{z}=\mathbf{u},\\
		& \qquad |\mathbf{z}(i)| = 1, i=1,2,...,NL.
	\end{aligned}
\end{equation}
where the reformulation process and the definition of $\mathcal{F}(\mathbf{u}, \mathbf{x})$ are detailed in Appendix \ref{PDD}.

\end{proposition}

Afterwards, we apply the augmented Lagrange method to incorporate the equality constraints into the objective function. Specifically, by introducing a set of Lagrange multipliers $\bm{\lambda} = \{\bm{\lambda}_1, \bm{\lambda}_2\}$ and penalty parameter $\xi$, $(P4)$ can be formulated as\vspace{-3mm}
\begin{equation}\vspace{-2mm}
\begin{aligned}
(P5) \quad & \min_{\mathbf{u}, \mathbf{x}, \mathbf{z}} \quad \mathcal{G}(\mathbf{u}, \mathbf{x}, \mathbf{z})\\
& s.t. \quad |\mathbf{z}(i)| = 1, \ i=1,2,...,NL,
\end{aligned}
\end{equation}
where\vspace{-2mm}
\begin{equation}\label{la2}\vspace{-4mm}
\mathcal{G}(\mathbf{u}, \mathbf{x}, \mathbf{z}) = \mathcal{F}(\mathbf{u}, \mathbf{x}) + \frac{1}{2\xi}(\|\mathbf{x}-\mathbf{u}+\xi \bm{\lambda}_1\|^2+\|\mathbf{z}-\mathbf{u}+\xi \bm{\lambda}_2\|^2).
\end{equation}

Note that the proposed PDD-based algorithm is a nested loop iterative algorithm, where an augmented Lagrange problem is solved in the inner loop and Lagrange multipliers or the penalty parameter are updated in the outer loop.

\subsubsection{Inner-loop Iterative Algorithm for Handling $(P6)$}
\
\newline
\indent $\bm{Step} \ \textit{1}:$ Optimizing $\mathbf{u}$ with fixed $\mathbf{x}$, $\mathbf{z}$ and $\bm{\lambda}$.

By omitting elements independent of $\mathbf{u}$, the resulting subproblem is given by\vspace{-4mm}
\begin{equation}\vspace{-5mm}
(P5-1) \quad \min_{\mathbf{u}} \mathcal{G}_1(\mathbf{u}) = \mathbf{u}^T \mathbf{R} \mathbf{u}^* + \mathbf{u}^T \mathbf{t} + \mathbf{t}^H \mathbf{u}^* + \frac{1}{2\xi}(\|\mathbf{x}-\mathbf{u}+\xi \bm{\lambda}_1\|^2+\|\mathbf{z}-\mathbf{u}+\xi \bm{\lambda}_2\|^2),
\end{equation}
{where the definitions of $\mathbf{R}$ and $\mathbf{t}$ are given in Appendix \ref{definition}.}

It can be shown that $\mathbf{R}$ is a positive definite matrix, hence $(P5-1)$ is then an unconstrained convex problem, as such, the optimal solution can be obtained by utilizing the first-order optimality condition. However, this will introduce matrix inversion operations which incur high computation complexity. Instead, we apply the subgradient method in \cite{PDS} as follow\vspace{-3mm}
\begin{equation}\label{u}\vspace{-3mm}
\mathbf{u}^{t+1}=\mathbf{u}^t-\alpha_1 \cdot \partial \mathcal{G}_1(\mathbf{u}) / \partial \mathbf{u} |_{\mathbf{u}=\mathbf{u}^t},
\end{equation}
where $t$ is the iteration index; $\alpha_1>0$ is the sufficiently small step size to guarantee the convergence; $\partial \mathcal{G}_1(\mathbf{u}) / \partial \mathbf{u} |_{\mathbf{u}=\mathbf{u}^t}$ is the subgradient of $\mathcal{G}_1(\mathbf{u})$ at $\mathbf{u}^t$, obtained by $\partial \mathcal{G}_1(\mathbf{x}) / \partial \mathbf{u} |_{\mathbf{u}=\mathbf{u}^t} = (\mathbf{R}+\frac{1}{\xi}\mathbf{I}_{LN}) \cdot(\mathbf{u}^t)^*+\mathbf{t}-\frac{1}{2\xi}(\mathbf{x}+\mathbf{z}+\xi \lambda_1+\xi \lambda_2)^*$.

$\bm{Step 2}:$ Optimizing $\mathbf{x}$ with fixed $\mathbf{u}$, $\mathbf{z}$ and $\bm{\lambda}$.

By omitting elements independent of $\mathbf{x}$, the resulting subproblem is given by\vspace{-2mm}
\begin{equation}\vspace{-2mm}
(P5-2) \quad \min_{\mathbf{x}} \mathcal{G}_2(\mathbf{x}) = \mathbf{x}^T \mathbf{Q} \mathbf{x}^* + \mathbf{x}^T \mathbf{a} + \mathbf{a}^H \mathbf{x}^* + \frac{1}{2\xi}\|\mathbf{x}-\mathbf{u}+\xi \bm{\lambda}_1\|^2,
\end{equation}
{where the definitions of $\mathbf{Q}$ and $\mathbf{a}$ are given in Appendix \ref{definition}.}

Again, it can be proved that the matrix $\mathbf{Q}$ is a positive definite matrix. Therefore, by using the subgradient method, we can get\vspace{-5mm}
\begin{equation}\label{x}\vspace{-5mm}
\mathbf{x}^{t+1}=\mathbf{x}^t-\alpha_1 \cdot \partial \mathcal{G}_2(\mathbf{x}) / \partial \mathbf{x} |_{\mathbf{x}=\mathbf{x}^t}
\end{equation}
where $\partial \mathcal{G}_2(\mathbf{x}) / \partial \mathbf{x} |_{\mathbf{x}=\mathbf{x}^t}=(\mathbf{Q}+\frac{1}{2\xi}\mathbf{I}_{LN})\cdot(\mathbf{x}^t)^*+\mathbf{a}-\frac{1}{2\xi}(\mathbf{u}-\xi \bm{\lambda}_1)$.

$\bm{Step 3}:$ Optimizing $\mathbf{z}$ with fixed $\mathbf{u}$, $\mathbf{x}$ and $\bm{\lambda}$.

By omitting elements independent of $\mathbf{z}$, the resulting subproblem is given by\vspace{-2mm}
\begin{equation}\vspace{-2mm}
\begin{aligned}
(P5-3)& \quad \min_{\mathbf{z}}  \frac{1}{2 \xi}\|\mathbf{z}-\mathbf{u}+\xi \bm{\lambda}_2\|^2\\
& \quad |\mathbf{z}(i)| = 1, \ i=1,2,...,NL.
\end{aligned}
\end{equation}

Because of the constraint: $|\mathbf{z}(i)|=1$, it can be verified that $\mathbf{z}=\angle(\mathbf{u}-\xi \bm{\lambda}_2)$ minimizes the Lagrangian function, where $\angle(\cdot)$ represents the phase vector of its argument.

\subsubsection{Outer-loop Penalty Parameters Update to Guarantee the Algorithm Convergence}
\
\newline
\indent We can update $\bm{\lambda}$ and $\rho$ in the outer loop to guarantee the convergence of the PDD-based algorithm according to the following equations:\vspace{-2mm}
\begin{equation}\vspace{-4mm}
\bm{\lambda}_1^{i+1} = \bm{\lambda}_1^i + \frac{\mathbf{x} - \mathbf{u}}{\xi}, \qquad
\bm{\lambda}_2^{i+1} = \bm{\lambda}_2^i + \frac{\mathbf{z} - \mathbf{u}}{\xi}, \qquad
\xi^{i+1} = c \xi^i.
\end{equation}
where $i$ denotes the iteration index.
\begin{remark}
It is worth noting that small $\xi$ makes the Lagrangian function dominated by the penalty terms, while large $\xi$ results in weak equality constraints. Considering these impacts, we can first initialize $\xi$ sufficiently large, and then decrease it gradually. Therefore, we set $c=0.7$ to achieve sufficiently good performance \cite{penalty_parameter}. Such an updating method aims to find appropriate penalty parameters to make the augmented Lagrangian method converge, and the details about the proof are referred to \cite{penalty_parameter,PDD}.
\end{remark}

\begin{remark}
Meanwhile, we can handle $(P5)$ by employing the block coordinate descent (BCD) approach. Specifically, we first split the optimization variables into three blocks: $\{\mathbf{u}\}$, $\{\mathbf{x},\ \mathbf{z}\}$, $\{\bm{\lambda_1},\ \bm{\lambda_2},\ \rho\}$, where variables of each block can be optimized in parallel. Thus, if a multi-core processor is available, it can reduce the computational time effectively by the BCD method.
\end{remark}

Ultimately, we summarize the principles of the RIS phase shifts design in \emph{Algorithm 1}, where the proposed PDD-based method acts as a subpart in \emph{Algorithm 2}.

\begin{algorithm}[h]
	\caption{The proposed algorithm of the phase shifts design at RISs}
	\begin{algorithmic}[1]
		\State Initialize the RIS phase shifts $\mathbf{u}^0$ to feasible values and set the iteration index $j=0$;
		\Repeat
		\State Update $r_k^{j+1} = \frac{\sum\limits_{s=1}^S \sqrt{\eta_{ks}}(\|\bar{\mathbf{h}}_{ks}\|^2+ M \chi_{ks})}{\mathcal{A}_k+\sum\limits_{i \neq k}^K \mathcal{B}_{k,i} + N_0}$ for $k=1,...,K$;
		\State Update $\kappa_k^{j+1} = \big{[}|r_k|^2(\mathcal{A}_k+\sum_{i \neq k}^K \mathcal{B}_{k,i})-2\text{Re}\{\sum\limits_{s=1}^S \sqrt{\eta_{ks}}(\|\bar{\mathbf{h}}_{ks}\|^2+ M \chi_{ks})\}+|r_k|^2N_0+1\big{]}^{-1}$ for $k=1,...,K$;
		\State Update $\mathbf{u}^{j+1}$ by using the Algorithm 2 and adjust $\bm{\Phi}$ accordingly;
		\State Set $j \leftarrow j+1$.
		\Until{the objective function of $(P2)$ converges or $j$ reaches the maximum iteration number }
	\end{algorithmic}
\end{algorithm}

\begin{algorithm}[h]
	\caption{PDD-based algorithm for minimizing $\mathcal{G}(\mathbf{u}, \mathbf{x}, \mathbf{z})$}
	\begin{algorithmic}[1]
		\State Initialize variables $\mathbf{u}^0, \mathbf{x}^0$, $\mathbf{z}^0$, $\bm{\lambda}$, $\xi$ and $\alpha_1$ to feasible values and set the outer-loop iteration index $i=0$;
		\Repeat
		\State Set the inner-loop iteration index $t=0$
		\Repeat
		\State Calculate the value of $\mathbf{R}$ and $\mathbf{t}$, and update $\mathbf{u}^{t+1}=\mathbf{u}^t-\alpha_1 \cdot \partial \mathcal{G}_1(\mathbf{u}) / \partial \mathbf{u} |_{\mathbf{u}=\mathbf{u}^t}$;
		\State Calculate the value of $\mathbf{Q}$ and $\mathbf{a}$, and update $\mathbf{x}^{t+1}=\mathbf{x}^t-\alpha_1 \cdot \partial \mathcal{G}_2(\mathbf{x}) / \partial \mathbf{x} |_{\mathbf{x}=\mathbf{x}^t}$ and $\mathbf{z}^{t+1}=\angle(\mathbf{u}^{t+1}-\xi \bm{\lambda}_2)$ in parallel;
		\State Set $j \leftarrow j+1$.
		\Until{ the value of $\mathcal{G}(\mathbf{u}, \mathbf{x}, \mathbf{z})$ converges or $j$ reaches the maximum iteration number }
		\State Update $\bm{\lambda}_1^{i+1} = \bm{\lambda}_1^i + \frac{\mathbf{x} - \mathbf{u}}{\xi}$, $\bm{\lambda}_2^{i+1} = \bm{\lambda}_2^i + \frac{\mathbf{z} - \mathbf{u}}{\xi}$ and $\xi^{i+1} = c \xi^i$ in parallel;
		\State Set $i \leftarrow i+1$
		\Until{ the max value of the equality constraint $\max \{ \|\mathbf{x}-\mathbf{u}\|^2, \|\mathbf{z}-\mathbf{u}\|^2 \}$ is below a threshold or $i$ reaches the maximum iteration number }
	\end{algorithmic}
\end{algorithm}
\vspace{-6mm}
\subsection{Power Allocation at the BSs}
The inner weighted sum-rate maximization problem can be written as\vspace{-2mm}
\begin{subequations}\label{OF2}
	\begin{equation}\vspace{-4mm}
		\begin{aligned}
			(P6) \quad \max_{\bar{\bm{\eta}}} \sum\limits_{k=1}^K \mu_k \log_2 \bigg{(} 1+\frac{ \| \sum\limits_{s=1}^S \sqrt{\eta_{ks}} \|\mathbf{h}_{ks}\|^2 \| ^2 }{  \sum\limits_{i \neq k}^K \sum\limits_{s=1}^S \sqrt{\eta_{is}} \|\mathbf{h}_{ks}^T \mathbf{h}_{is}^{*} \|^2 +N_0} \bigg{)}
		\end{aligned}
	\end{equation}
	\begin{equation}\vspace{-3mm}
		s.t. \quad  \sum\limits_{k=1}^K \eta_{sk} \|\mathbf{h}_{sk}\|^2 \leq P_{s,\max} , \quad s = \{1,...,S\}.
	\end{equation}
\end{subequations}

Since the objective function of $(P6)$ is a non-convex problem w.r.t. $\bar{\bm{\eta}}$, the optimal solution is intractable. We first apply the Lagrangian dual transform proposed in \cite{fractional}, and the objective function can be equivalently written as\vspace{-4mm}
\begin{equation}\label{epsilon_k}\vspace{-2mm}
	\sum_{k=1}^K\mu_k\log_2(1+\epsilon_k)-\sum_{k=1}^K\mu_k\epsilon_k+\sum_{k=1}^K\frac{(1+\epsilon_k)\mu_k\gamma_k}{1+\gamma_k}, \quad k \in \{1,...,K\},
\end{equation}
where $\gamma_k$ is the SINR of the $k$-th UE and $\epsilon_k$ is an auxiliary variable. The optimal $\epsilon_k$ can be obtained by setting the derivative to zero, and we get $\epsilon_k=\gamma_k$. Then, for the fixed optimal $\epsilon_k$, by discarding irrelevant parts to $\gamma_k$ in (\ref{epsilon_k}), problem $(P6)$ reduces to\vspace{-3mm}
\begin{equation}\vspace{-4mm}
	\begin{aligned}
		(P6') \quad & \max_{\bar{\bm{\eta}}} \sum\limits_{k=1}^K \frac{ (1+\epsilon_k) \mu_k \| \sum\limits_{s=1}^S \sqrt{\eta_{ks}} \|\mathbf{h}_{ks}\|^2 \|^2 }{  \sum\limits_{i=1}^K \|\sum\limits_{s=1}^S \sqrt{\eta_{is}} \mathbf{h}_{ks}^T \mathbf{h}_{is}^{*} \|^2 +N_0} \\
		& s.t. \quad (\ref{OF2}b).
	\end{aligned}
\end{equation}

Since $(P6')$ is a multi-ratio fractional programming problem, the quadratic transform in \cite{fractional} can be used. As such, the objective function can be rewritten as\vspace{-3mm}
\begin{equation}\label{OF3}\vspace{-3mm}
	\sum_{k=1}^K \bigg{\{} 2 \sqrt{\mu_k(1+\epsilon_k)} \text{Re}\{ y_k^* \sum_{s=1}^S \sqrt{\eta_{ks}} \|\mathbf{h}_{ks}\|^2 \} -|y_k|^2(\sum_{i=1}^K \|\sum_{s=1}^S \sqrt{\eta_{is}} \mathbf{h}_{ks}^T \mathbf{h}_{is}^{*} \|^2 +N_0) \bigg{\}},
\end{equation}
where $y_k = \frac{\sqrt{\mu_k(1+\epsilon_k)}\sum\limits_{s=1}^S \sqrt{\eta_{ks}} \|\mathbf{h}_{ks}\|^2}{\sum\limits_{i=1}^K \|\sum\limits_{s=1}^S \sqrt{\eta_{is}} \mathbf{h}_{ks}^T \mathbf{h}_{is}^{*} \|^2 +N_0}$. Then, the problem $(P6')$ can be reformulated as\vspace{-4mm}
\begin{equation}
	\begin{aligned}
		(P7)\quad &\min_{\bar{\bm{\eta}}} \ f(\bar{\bm{\eta}})= \bar{\bm{\eta}} \bm{\Xi} \bar{\bm{\eta}}^H - 2\text{Re} \{\bar{\bm{\eta}} \bm{\varpi}^T \} + \delta \\
		& s.t. \  \bar{\bm{\eta}} \bm{\Pi}_s \bar{\bm{\eta}}^H \leq P_{s,\max},
	\end{aligned}
\end{equation}
where $\bm{\Xi}= \text{diag}\{ \mathbf{D}_{1},...,\mathbf{D}_K \} \in \mathbb{C}^{KS \times KS}$, $\mathbf{D}_i = \sum_{k=1}^K |y_k|^2 \mathbf{d}_{ki} \mathbf{d}_{ki}^H$, $\mathbf{d}_{ki} = [\mathbf{h}_{k1}^T \mathbf{h}_{i1}^*,...,\mathbf{h}_{kS}^T \mathbf{h}_{iS}^*]^T$, $\bm{\varpi}=[\sqrt{\mu_1(1+\epsilon_1)}y_1^*\mathbf{c}_1^T,...,\sqrt{\mu_K(1+\epsilon_K)}y_K^*\mathbf{c}_K^T]$, $\mathbf{c}_k = [\|\mathbf{h}_{k1}\|^2,...,\|\mathbf{h}_{kS}\|^2]^T$, $\delta=\sum_{k=1}^K|y_k|^2 N_0$, $\bm{\Pi}_s=\mathbf{E}_s \otimes (\mathbf{i}_s^H \mathbf{i}_s)$, $\mathbf{E}_s = \text{diag}\{\|\mathbf{h}_{1s}\|^2,...,\|\mathbf{h}_{Ks}\|^2\}$, and $\mathbf{i}_s$ is an elementary vector with a one at the $s$-th position and zeros at other positions.

Since $\bm{\Xi}$ is a positive semidefinite matrix, the resulting objective function of $(P7)$ is a convex function w.r.t. $\bar{\bm{\eta}}$. Hence, the optimization problem $(P7)$ is a quadratic constrained quadratic programming (QCQP), which can be solved by many existing methods such as alternating direction method of multipliers (ADMM) \cite{ADMM}. However, the adoption of ADMM method will inevitably operate matrix inversions which cause high computation complexity. Therefore, we use the PDS method \cite{PDS} to obtain an effective solution free of matrix inversion operations. Following the PDS methodology, the augmented Lagrangian function is given as\vspace{-3mm}
\begin{equation}\label{la1}
	\mathcal{L}(\bar{\bm{\eta}}, \bm{\zeta})=f(\bar{\bm{\eta}})+\bm{\zeta}^T \mathbf{g}(\bar{\bm{\eta}})+\frac{\rho}{2}\|\mathbf{g}(\bar{\bm{\eta}})\|^2,
\end{equation}
where $\bm{\zeta} \in \mathbb{C}^{S \times 1}$ is the Lagrange multiplier vector; $\rho > 0$ denotes the penalty parameter; $\mathbf{g}(\bar{\bm{\eta}})=[{g}_1^+(\bar{\bm{\eta}}),...,{g}_S^+(\bar{\bm{\eta}})]$, and $g_s^+(\bar{\bm{\eta}})=\max (0,\ \bar{\bm{\eta}} \bm{\Pi}_{s} \bar{\bm{\eta}}^H-P_{s,\max})$.

Then, the $j$-th iteration of the PDS-based algorithm can be described as follows:\vspace{-3mm}
\begin{equation}\label{pds}
	\begin{bmatrix}
		\bar{\bm{\eta}}^{j+1}, \
		\bm{\zeta}^{j+1}
	\end{bmatrix}
	=
	\begin{bmatrix}
		\bar{\bm{\eta}}^{j}, \
		\bm{\zeta}^{j}
	\end{bmatrix}
	-\alpha_2 \cdot
	\begin{bmatrix}
		\partial \mathcal{L}/\partial \bar{\bm{\eta}}|_{\bar{\bm{\eta}}=\bar{\bm{\eta}}^j}, \
		-\partial \mathcal{L}/\partial \bm{\zeta}|_{\bm{\zeta}=\bm{\zeta}^j}
	\end{bmatrix},
\end{equation}
where $\alpha_2>0$ denotes the sufficiently small step length to guarantee the convergence, $\partial \mathcal{L}/\partial \bar{\bm{\eta}} = \bar{\bm{\eta}}^* \bm{\Xi}^T - \bm{\varpi}+\sum_{s=1}^S(\zeta_s+\rho g_s^+(\bar{\bm{\eta}}))\bm{\omega}_s(\bar{\bm{\eta}})$ and $\partial \mathcal{L}/\partial \bm{\zeta} = \mathbf{g}(\bar{\bm{\eta}})$, wherein the auxiliary function $\bm{\omega}_s(\bm{\eta})$ can be written as
\begin{equation}
	\bm{\omega}_s(\bar{\bm{\eta}})=\left\{\begin{matrix}
		\bar{\bm{\eta}}^* \bm{\Pi}_s^T, \qquad \mathbf{g}_s(\bm{\eta})>0\\
		\mathbf{0}_{KS}, \qquad \mathbf{g}_s(\bm{\eta}) \leq 0.
	\end{matrix}\right.
\end{equation}

In conclusion, the proposed PDS-based algorithm for solving the power allocation optimization problem can be summarized in \emph{Algorithm 3}.

\begin{algorithm}[h]
	\caption{The PDS-based algorithm of the power allocation at BSs}
	\begin{algorithmic}[1]
		\State Initialize $\bm{\eta}$, $\bm{\zeta}$, $\rho$ and $\alpha_2$ to feasible values; calculate the value of the SINR $\bm{\gamma}^0$ and set the iteration index $j=0$;
		\Repeat
		\State Update $\epsilon_k=\gamma_k $ for $k \in \{1,...,K\}$;
		\State Update $y_k = \frac{\sqrt{\mu_k(1+\epsilon_k)}\sum\limits_{s=1}^S \sqrt{\eta_{ks}} \|\mathbf{h}_{ks}\|^2}{\sum\limits_{i=1}^K \|\sum\limits_{s=1}^S \sqrt{\eta_{is}} \mathbf{h}_{ks}^T \mathbf{h}_{is}^{*} \|^2 +N_0}$ for $k \in \{1,...,K\}$;
		\State Calculate the value of $\bm{\Xi}$, $\bm{\varpi}$ and $\bm{\Pi}$, and update $\bar{\bm{\eta}}$ and $\bm{\zeta}$ in Eq. (\ref{pds}) parallelly;
		\State Set $j \leftarrow j+1$.
		\Until{ the value of the Eq. (\ref{OF2}a) converges or $j$ reaches the maximum iteration number. }
	\end{algorithmic}
\end{algorithm}
\vspace{-5mm}
\subsection{Property Analysis of the Overall Algorithm}
In the alternating optimization framework, the passive beamformers of RISs are first optimized by \emph{Algorithm 1} with fixed BSs power allocation. Then with fixed RISs designs, the power-control coefficients of BSs are optimized by \emph{Algorithm 3}. Let $\mathcal{R}$ denote the objective function of the ergodic weighted sum-rate maximization problem $(P1)$. The convergence behavior and computational complexity of the overall algorithm are elaborated in the sequel.
\subsubsection{Convergence}
{In the $(i+1)$-th iteration of the overall algorithm, by performing the optimizations on RISs and BSs by \emph{Algorithm 1} and \emph{Algorithm 3}, the value difference between $\mathcal{R}(\bm{\Phi}^{i+1},\bar{\bm{\eta}}^{i+1})$ and $\mathcal{R}(\bm{\Phi}^{i},\bar{\bm{\eta}}^{i})$ is explored as follows, where $\mathcal{R}(\bm{\Phi},\bar{\bm{\eta}})$ is the achievable sum-rate with $\bm{\Phi}$ being the RISs phase shifts and $\bar{\bm{\eta}}$ being the BSs power allocation coefficients.}
\begin{enumerate}[(a)]
	\item {In \emph{Algorithm 1}, the RISs optimization problem is first transformed to the $\mathcal{G}(\mathbf{u},\mathbf{x},\mathbf{z})$ minimization problem equivalently by the novel ergodic WMMSE reformulation. Then, the PDD-based method is developed to solve it, and it is a nested loop iteration algorithm. Specifically, in the inner-loop iteration for updating $\mathbf{u}$, $\mathbf{x}$ and $\mathbf{z}$, the objective function $\mathcal{G}(\mathbf{u},\mathbf{x},\mathbf{z})$ is guaranteed to decrease monotonously, since the subprolems all belong to the convex optimizations, i.e,\vspace{-3mm}
	\begin{equation}\vspace{-3mm}
	\mathcal{G}^{t+1}(\mathbf{u},\mathbf{x},\mathbf{z}) \leq \mathcal{G}^{t}(\mathbf{u},\mathbf{x},\mathbf{z}).
	\end{equation}
	where $t$ denotes the number of inner iterations of the \emph{Algorithm 2}. On the other hand, the updatings of $\bm{\lambda}$ and $\rho$ make the augumented Lagrangian method converge \cite{PDD,penalty_parameter} as stated in the \emph{Remark 1} in the manuscript. Thus, the \emph{Algorithm 1} is guaranteed to increase $\mathcal{R}(\bm{\Phi}^{i},\bar{\bm{\eta}}^{i})$ and converge, and we have\vspace{-3mm}
	\begin{equation}\vspace{-3mm}
	\mathcal{R}(\bm{\Phi}^{i+1},\bar{\bm{\eta}}^{i}) \geq \mathcal{R}(\bm{\Phi}^{i},\bar{\bm{\eta}}^{i}),
	\end{equation}}
	\vspace{-5mm}
	\item {In \emph{Algorithm 3}, the BSs power allocation optimization problem is transformed to a QCQP. Then, the PDS-based method is proposed to minmize $\mathcal{L}(\bm{\eta},\bm{\zeta})$. In the $t+1$-th iteration of \emph{Algorithm 3}, the set-valued mapping is based on the Karush-Kuhn-Tucker (KKT) operation associated with $\mathcal{L}(\bm{\eta},\bm{\zeta})$, and have\vspace{-3mm}
	\begin{equation}\vspace{-3mm}
	\mathcal{L}(\bm{\eta}^{t+1},\bm{\zeta}^{t+1}) \leq \mathcal{L}(\bm{\eta}^{t+1},\bm{\zeta}^{t}) \leq \mathcal{L}(\bm{\eta}^{t},\bm{\zeta}^{t}).
	\end{equation}
	Hence, for a constant step size $\alpha_2$, the algorithm is guaranteed to converge around the optimal value, i.e.,\vspace{-3mm}
	\begin{equation}\vspace{-3mm}
	\lim\limits_{t \to \infty} \mathcal{L}(\bm{\eta}^{t+1},\bm{\zeta}^{t+1}) - \mathcal{L}^*(\bm{\eta},\bm{\zeta}) < \epsilon,
	\end{equation}
	where $\mathcal{L}^*(\bm{\eta},\bm{\zeta})$	denotes the optimal value of the $\mathcal{L}(\bm{\eta},\bm{\zeta})$, and the gap value $\epsilon$ is a function of parameter $\alpha_2$, and decreases with it. Therefore, adopting a sufficiently small step length can facilitate the convergence \cite{PDS}. Thus, the \emph{Algorithm 3} is guaranteed to increase $\mathcal{R}(\bm{\Phi}^{i+1},\bar{\bm{\eta}}^{i})$ and converge, and have\vspace{-3mm}
	\begin{equation}\vspace{-3mm}
	\mathcal{R}(\bm{\Phi}^{i+1},\bar{\bm{\eta}}^{i+1}) \geq \mathcal{R}(\bm{\Phi}^{i+1},\bar{\bm{\eta}}^{i}).
	\end{equation}}
\end{enumerate}
\vspace{-5mm}
{Summarizing the above proof, the processings of updating BS power-control coefficients and RIS phase shifts are all monotonously increasing, and we can\vspace{-3mm}
\begin{equation}\vspace{-3mm}
\mathcal{R}(\bm{\Phi}^{i},\bar{\bm{\eta}}^{i}) \geq \mathcal{R}(\bm{\Phi}^{i-1},\bar{\bm{\eta}}^{i-1}),
\end{equation}
which indicates $\mathcal{R}$ is non-decreasing after each iteration. Besides, since $\mathcal{R}$ has a limited upper-bound, the proposed algorithm is graranteed to converge.}

\subsubsection{Computational Complexity}
We then analyze the computational complexity of two subproblems, separately.
\begin{enumerate}[(a)]
\item \emph{RISs design :} The complexity of \emph{Algorithm 1} for updating $\mathbf{r}$ and $\bm{\kappa}$ are on the order of $\mathcal{O}(M^2)$ and $\mathcal{O}(MLN)$, respectively. Then, in the inner-loop iteration of \emph{Algorithm 2}, the complexity for solving three subproblems are all on the order of $\mathcal{O}(L^2N^2)$. Therefore, regarding the dominant terms, the overall complexity for updating $\bm{\Phi}$ is $\mathcal{O}(I_1 I_2 (M^2+MLN+L^2N^2))$, where $I_1$ and $I_2$ are the inner-loop and outer-loop iteration numbers, respectively.

\item \emph{BSs power-control coefficients design:} In each iteration of \emph{Algorithm 1}, the complexity for updating $\bm{\epsilon}$ and $\mathbf{y}$ are both on the order of $\mathcal{O}(M^2)$. Then, the complexity of solving $(P7)$ is on the order of $\mathcal{O}(K^2S^2)$ thanks to applying the PDS method to deal with the QCQP. To summarize, the complexity for updating $\bar{\bm{\eta}}$ is on the order of $\mathcal{O}(I_3(M^2+K^2S^2))$, where $I_3$ denotes the maximum iteration number.

\end{enumerate}
{The joint beamformers at BSs and RISs are in the iteration framework, so the overall computational complexity is on the order of $\mathcal{O}\bigg(I_4 \cdot\big( I_1 I_2 (M^2+MLN+L^2N^2)+ I_3(M^2+K^2S^2) \big) \bigg)$, where $I_4$ is the iteration number of the proposed algorithm.} 

{Since the number of RISs elements dominates the computational complexity, it can be regarded as $\mathcal{O}(L^2N^2)$. As a benchmark, the QCQP is commonly solved by the ADMM method \cite{cellfree-ris3,ADMM1} with the complexity of $\mathcal{O}(L^3N^3)$, which is higher than that of the proposed algorithm.}

\vspace{-3mm}
\section{NUMERICAL RESULTS}\label{simulation}
In this section, we provide numerical results to validate the theoretical achievable sum-rate expression, evaluate the performance of the proposed algorithm and draw useful insights. We simulate a RIS-assisted cell-free network where BSs, RISs and UEs are randomly distributed over a square dense urban area of size $100 \times 100$ $m^2$. The distance-dependent large scale path-loss coefficient is modeled as $\beta(d)=C_0 (\frac{d}{d_0})^{-\alpha}$ \cite{largescale}, where $C_0$ is the path loss exponents at the reference distance $d_0=1$m, $d$ represents the individual link distance and $\alpha$ denotes the path loss exponent. {Unless otherwise specified, the setup given by Table \ref{tab1} is used.}
\begin{table}[ht]
	\renewcommand\arraystretch{1.3}
	\begin{center}
		\caption{Parameters used in simulations}
		\vspace{3mm}
		\label{tab1}
		\begin{tabular}{ | m{4.7cm} <{\centering}| m{4.8cm}<{\centering} |}
			\hline 
			\textbf{Parameters} &\textbf{Values}  \\  \hline
			Path loss & $C_0 = -30$ dB \\ \hline
			Path loss exponent & $\alpha_D=3.5$, $\alpha_{BR}=2.2$ , $\alpha_{RU}=2.8$ \cite{pathloss} \\ \hline
			Noise density power & $N_0 = -80$ dBm \\ \hline
			Transmit power & $P_{s.\max} = 10$ dBm  \\ \hline
			Rician $K$ factor &   $K_{ls} = K_{r,lk} = K_{d,ks} = 3+\sqrt{12}$  \cite{Kfactor}  \\ \hline
			Number of BSs &   $S=3$    \\ \hline
			Number of antennas per BS &   $M=4$    \\ \hline
			Number of RISs &   $L=3$    \\ \hline
			Number of elements per RIS & $N=64$ \\ \hline
			Number of UEs & $K=4$ \\ \hline
			Priority weight of UEs & $\mu_k = 1$, $\forall k$ \\ \hline
			
		\end{tabular}
	\end{center}
\end{table}

\subsection{Validation of the Achievable Weighted Sum-rate Expression}
\begin{figure}[ht]
	\centerline{\includegraphics[width=0.48\textwidth]{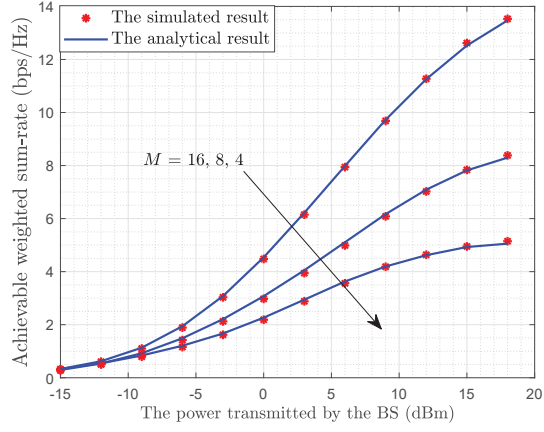}}  
	\caption{Achievable weighted sum-rate of the system with different $M$.}\label{fig1}
\end{figure}

Fig. \ref{fig1} validates the analytical expression of the achievable weighted sum-rate given in \emph{Theorem 1}, where we randomly generate RISs phase shifts and consider equal receiving power at all users, i.e., $\eta_{ks}=P_{s,\max}/(K \|\mathbf{h}_{ks}\|^2)$, $\forall k$. As can be readily observed, the analytical and simulation curves almost overlap with each other under different $M$ scenarios, thereby validating the accuracy of the provided analytical expression. It can also be observed that the sum-rate increases with $M$, which means that a larger number of BS antennas can provide higher spatial diversity gain.
\vspace{-2mm}
\subsection{Impact of the Number of RIS elements with no LoS components}
\begin{figure}[ht]
	\centerline{\includegraphics[width=0.48\textwidth]{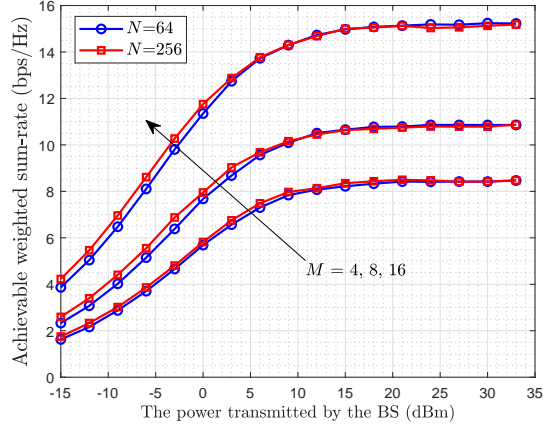}}  \vspace{-2mm}
	\caption{The impact of $N$ on the achievable weighted sum-rate.}\label{fig2}
\end{figure}
\vspace{-2mm}
Fig. \ref{fig2} illustrates the impact of $N$ on the achievable weighted sum-rate without LoS components. As predicted by \emph{Corollary 1}, in a low SNR regime, increasing the number of RIS elements or BS antennas both can improve the system performance, while in a high SNR regime, the achievable weighted sum-rate is not affected by the number of RIS elements. This is because the system performance has a power gain on the order of $\mathcal{O}(M)$ in a high SNR regime with no LoS components.

\subsection{Evaluation of the Proposed Algorithm}\vspace{-1mm}
\begin{figure}[ht]
	\centerline{\includegraphics[width=0.48\textwidth]{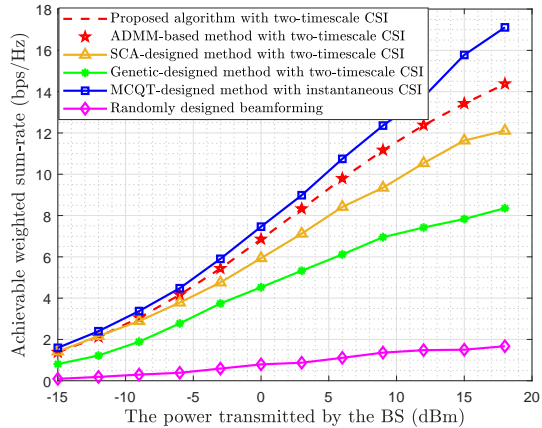}}  \vspace{-2mm}
	\caption{Comparison between the proposed scheme and benchmarks.}\label{fig3}
\end{figure}
\vspace{-3mm}
{Fig. \ref{fig3} compares our proposed algorithm with conventional benchmarks, i.e., ADMM-based RIS phase shifts scheme with two-timescale CSI\cite{S-CSI1}, successive convex approximation (SCA)-designed RISs with two-timescale CSI \cite{tts}, genetic-designed RISs and uniform power allocation with two-timescale CSI \cite{genetic}, multidimensional complex quadratic transform (MCQT)-based RISs with instantaneous CSI \cite{cellfree-ris1} and random beamforming design. Except for the random beamforming scheme, all algorithms are assumed to adopt MR precoding at BSs.}

{It can be observed in Fig. \ref{fig3} that the MCQT-designed scheme with instantaneous CSI achieves the best performance. This is reasonable, because instantaneous CSI can be used to control RISs to reflect beams aimed at UEs at any time. It is also shown that the proposed method achieves the comparable performance with the ADMM-based RIS phase shifts scheme.} Comparing the schemes with two-timescale CSI, we see that the performance of our proposed algorithm outperforms that of the remaining methods, which demonstrates the satisfactory performance of the proposed algorithm. In particular, the lack of a tractable achievable sum-rate closed-form expression in the SCA-designed method will inevitably incur performance loss. Besides, the utilization of heuristic optimization on RISs and uniform power-control scheme in a genetic-designed method would impair the performance distinctly.

\begin{figure}[ht]
	\centerline{\includegraphics[width=0.48\textwidth]{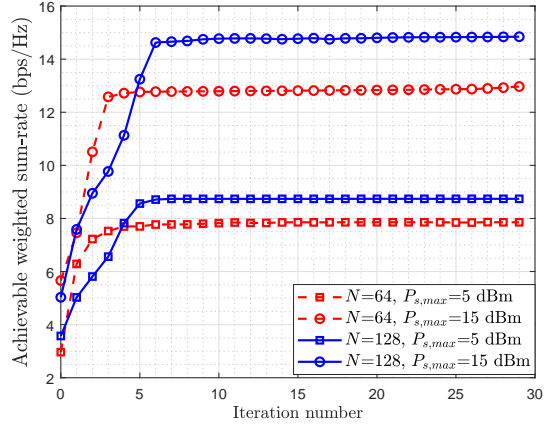}}  \vspace{-3mm}
	\caption{Weighted sum-rate versus the number of iterations.}\label{fig4}
\end{figure}

Fig. \ref{fig4} plots the convergence behavior of the weighted sum-rate with different $N$ and $P_{s,max}$. As can be readily observed, the weighted sum-rate increases rapidly and converges after few iterations. It can also be observed that the system with more RIS elements requires more iterations to converge, which indicates that the number of the RIS elements will have a great impact on the computational complexity of the proposed algorithm.

\subsection{Performance Gain Obtained by the Cell-free Paradigm}\vspace{-2mm}
\begin{figure}[ht]
	\centerline{\includegraphics[width=0.48\textwidth]{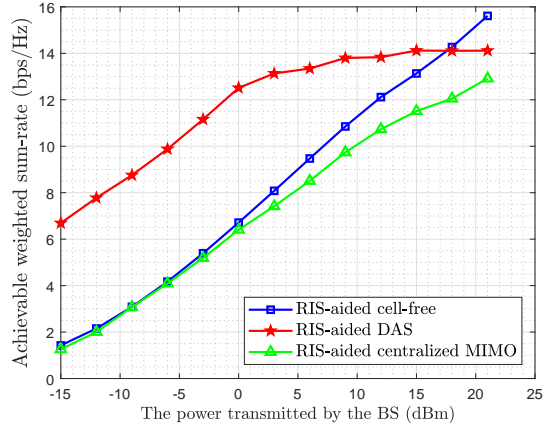}}  \vspace{-2mm}
	\caption{Comparison between the cell-free paradigm and benchmarks.}\label{fig5}
\end{figure}
Fig. \ref{fig5} compares the cell-free paradigm with other benchmarks, i.e., the distributed antenna system (DAS) \cite{das} and centralized MIMO system \cite{centralized}. Specifically, multiple remote antennas are deployed distributedly and connected to each BS in the DAS system, while all antennas are integrated into one BS in the centralized MIMO system. For a fair comparison, the same number of transmit antennas and same transmit power is assumed. It can be observed that, in the low SNR regime, the DAS system performs much better than the other schemes. This is because distributed antennas are more likely to be deployed close to UEs, thus gaining more spatial diversity. However, as the transmit power increases, the inter-user interference will become more severe in the distribution paradigm, and thus the cell-free system outperforms the DAS system at high SNR. This is a rather encouraging result, since the cell-free paradigm can achieve better performance with less channel estimation overhead. Moreover, the performance of the centralized MIMO system is inferior to the other schemes, since the centralized setting would lose the spatial diversity gain.
\subsection{Performance Gain Obtained by RISs}
\begin{figure}[http]
	\centering
	\subfigure[Location of the BSs, UEs and RISs.]{
		\label{fix_location}
		\includegraphics[width=0.48\textwidth]{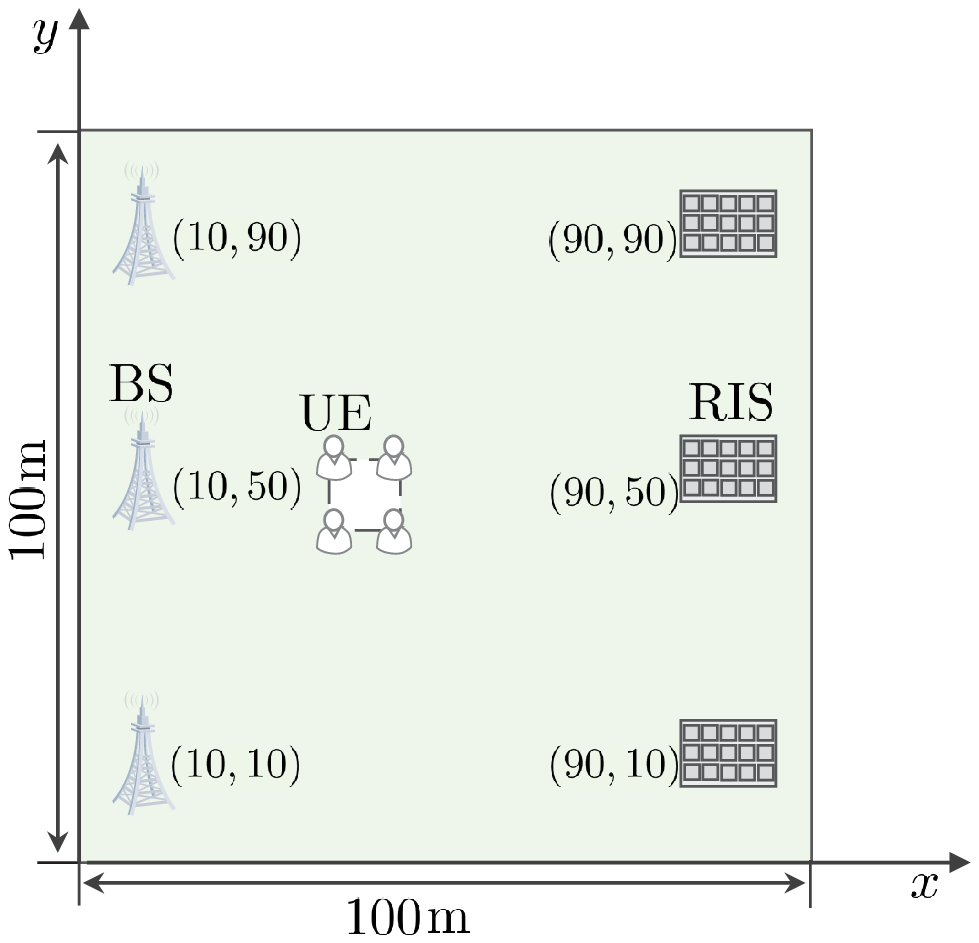}}
	\subfigure[Weighted sum-rate versus the location of UEs.]{
		\label{location}
		\includegraphics[width=0.48\textwidth]{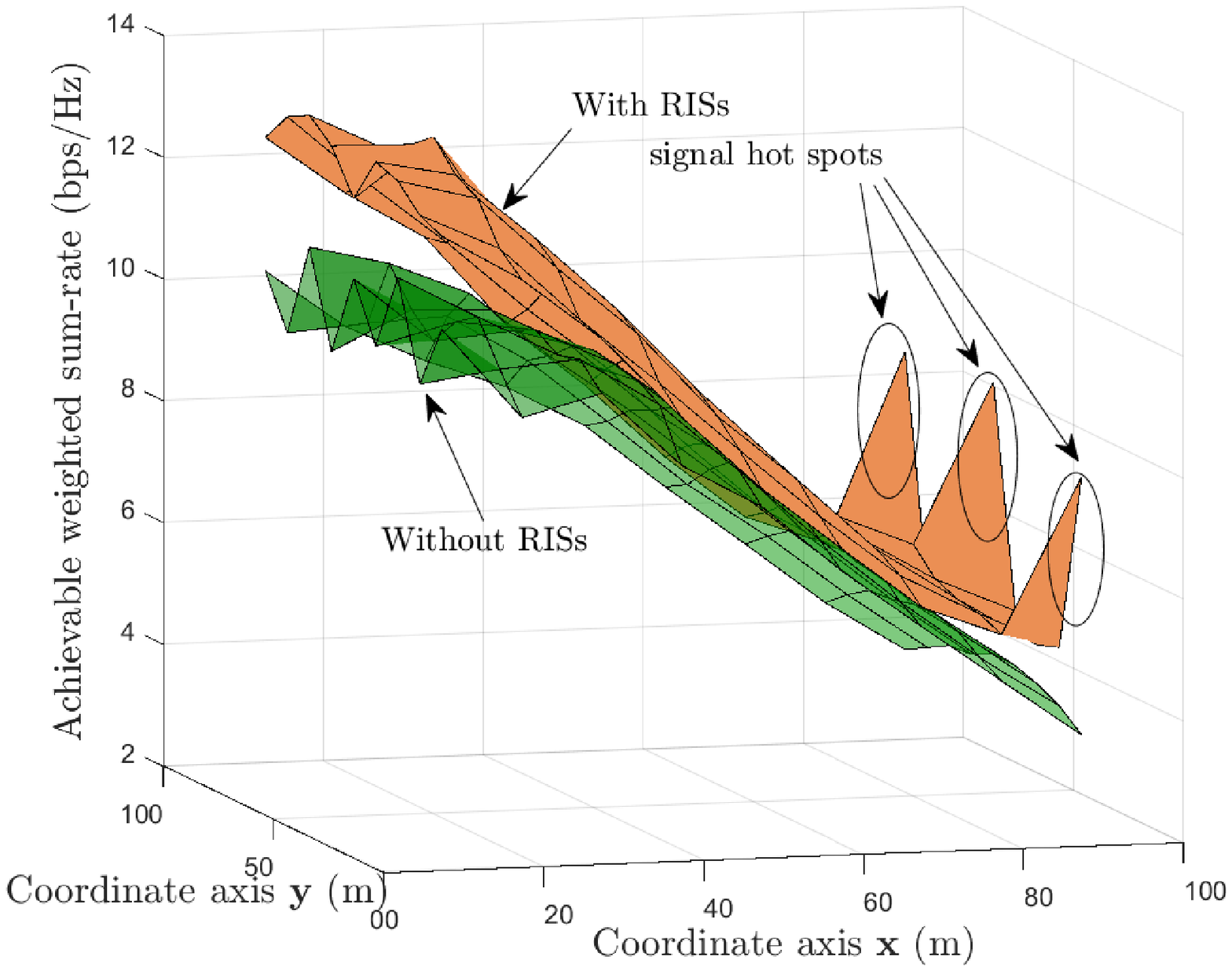}}
	\caption{Comparison between the scenarios with RISs and without RISs.}
	\label{location_fig}
\end{figure}
\vspace{-3mm}
Fig.~\ref{location_fig} illustrates the effects of the deployment of RISs on the system performance. For the scenario with RISs, Fig.~\ref{fix_location} depicts the location of BSs, UEs and RISs, where four UEs are deployed at $(x-2,y-2)$, $(x-2,y+2)$, $(x+2,y-2)$ and $(x+2,y+2)$, respectively. For the scenario without RISs, the simulation setup just removes the RISs in Fig.~\ref{fix_location}. As shown in Fig.~\ref{location}, we plot the average weighted sum-rate versus the location of UEs in two scenarios, which can demonstrate the performance gain obtained by RISs. As can be readily observed, the performance of the scenario with RISs is always superior to that without RISs, especially when UEs are in the vicinity of BSs or RISs. For the case without RISs, the weighted sum-rate is a monotonically decreasing function when the relative distance between UEs and BSs becomes larger. Instead, the case with RISs does not show a similar trend, where local peak values appear in the vicinity of RISs, namely, signal hot spots are created nearby.

\vspace{-2mm}
\subsection{The Impact Evaluation of the Rician $K$ Factor}\vspace{-2mm}
\begin{figure}[ht]
	\centerline{\includegraphics[width=0.47\textwidth]{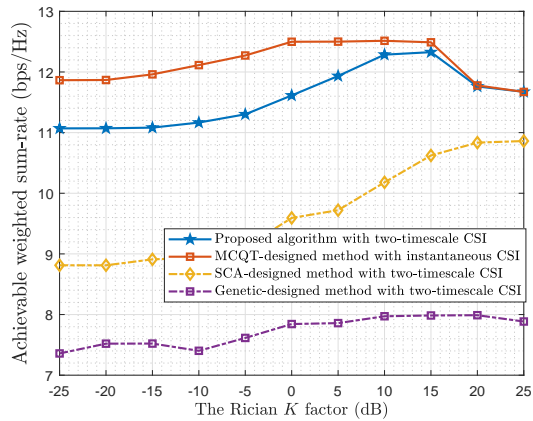}}  \vspace{-3mm}
	\caption{The impact of Rician $K$ factor on the weighted sum-rate.}\label{fig6}
\end{figure}
In Fig.~\ref{fig6}, we present the impact of Rician $K$ factor on the system performance versus different methods, where we assume $K_{ls}=K_{r,lk}=K_{d,ks}=K_{factor}$ for simplicity. It can be observed that the performance of the instantaneous CSI-based RIS method will first increase, then stay stable, and finally decrease with $K_{factor}$. For a small $K$ factor regime, the increase of $K_{factor}$ can facilitate the maximum beamforming gain. However, for a high $K$ factor regime, overly concentrated channels will lead to low spatial diversity and high inter-user interference, thus resulting in a low weighted sum-rate. Similar trends are observable in the performance of the proposed algorithm with two-timescale CSI. And as expected, the performance gap between them gradually narrows to zero with the increase of the Rician $K$ factor. This is because the CSI difference between these two algorithms is marginal for the high $K$ factor. However, there would be a constant performance gap between the SCA-based method and our proposed method. The reason is that the SCA algorithm may lead to a local optimum solution and result in performance loss.

\section{CONCLUSIONS}\label{conclusion}
In this paper, a closed-form analytical expression for the achievable sum-rate was first derived, which enables the evaluation of the impact of key parameters on system performance. To obtain more insights, a special case without LoS components was further investigated, where a power gain on the order of $\mathcal{O}(M)$ is achieved. Then, a low-complexity algorithm via the two-timescale transmission protocol was proposed in this paper, where the joint beamformers at BSs and RISs were facilitated via an alternating optimization framework to maximize the achievable weighted sum-rate. Specifically, the RISs beamformers were optimized based on the PDD method exploiting the statistical CSI, while the BSs beamformers were designed by the PDS method dependent on the instantaneous CSI. Besides, we proved the convergence of the proposed two-timescale beamforming algorithm and calculated its complexity. Numerical results validated the tightness of our derived analytical expression and showed the performance of the proposed algorithm. The impact of key system parameters such as the number of RIS elements, CSI settings and Rician factor was also evaluated. Moreover, the remarkable advantages from the adoption of the cell-free paradigm and the deployment of RISs exploitation were demonstrated intuitively.

\appendices
\section{The Expression of the Achievable sum-rate $C_k$}\label{ergodic_capacity}
The achievable sum-rate can be obtained approximately as in \cite{S-CSI1,approximation}:
\begin{equation}
\log_2 \bigg{(} 1+\frac{\mathbb{E}\{\| \sum\limits_{s=1}^S \sqrt{\eta_{ks}} \|\mathbf{h}_{ks}\|^2 \| ^2\}}{\sum\limits_{i \neq k}^K \mathbb{E}\{\|\sum\limits_{s=1}^S \sqrt{\eta_{is}} \mathbf{h}_{ks}^T \mathbf{h}_{is}^{*} \|^2\}+N_0} \bigg{)}
\end{equation}

\subsection{Calculate $\mathbb{E} \{\| \sum_{s=1}^S \sqrt{\eta_{ks}} \|\mathbf{h}_{ks}\|^2 \| ^2 \}$:}
We decompose $\mathbf{h}_{k}$ as $\mathbf{h}_{ks}=\bar{\mathbf{h}}_{ks}+\tilde{\mathbf{h}}_{ks}$ where $\bar{\mathbf{h}}_{ks}=a_{d,ks}\bar{\mathbf{h}}_{\mathrm{d},ks}+\sum_{l=1}^{L}a_{\mathrm{r},l k}a_{ls}\bar{\mathbf{G}}_{ls}^{T}\bm{\Theta}_{l}^{T}\bar {\mathbf{h}}_{\mathrm{r},lk}$, $\tilde{\mathbf{h}}_{ks}=\sum_{n=1}^{4}\mathbf{x}_{\mathrm{n},ks}$, $\mathbf{x}_{1,ks} = b_{\mathrm{d},ks}\tilde{\mathbf{h}}_{\mathrm{d},ks}$, $\mathbf{x}_{2,ks} = \sum_{l=1}^{L}{a_{\mathrm{r},l k}b_{ls}\tilde{\mathbf{G}}_{ls}^{T}\bm{\Theta}_{l}^{T}\bar{\mathbf{h}}_{\mathrm{r},l k}}$, $\mathbf{x}_{3,ks}= \sum_{l=1}^{L}{b_{\mathrm{r},l k}a_{ls}\bar{\mathbf{G}}_{ls}^{T}\bm{\Theta}_{l}^{T}\tilde{\mathbf{h}}_{\mathrm{r},l k}}$, $\mathbf{x}_{4,ks}=\sum_{l=1}^{L}{b_{\mathrm{r},l k}b_{ls}\tilde{\mathbf{G}}_{ls}^{T}\bm{\Theta}_{l}^{T}\tilde{\mathbf{h}}_{\mathrm{r},l k}}$, $a_{\mathrm{d}, ks} = \sqrt{\frac{\beta_{\mathrm{d}, ks} K_{\mathrm{d}, ks}}{K_{\mathrm{d}, ks}+1}}, b_{\mathrm{d}, ks}= \sqrt{\frac{\beta_{\mathrm{d}, ks}}{K_{\mathrm{d}, ks}+1}}, a_{\mathrm{r}, l k} = \sqrt{\frac{\beta_{\mathrm{r}, l k} K_{\mathrm{r}, l k}}{K_{\mathrm{r}, l k}+1}}, b_{\mathrm{r}, l k} = \sqrt{\frac{\beta_{\mathrm{r}, l k}}{K_{\mathrm{r}, l k}+1}}, a_{ls} = \sqrt{\frac{\beta_{ls} K_{ls}}{K_{ls}+1}}, b_{ls} = \sqrt{\frac{\beta_{ls}}{K_{ls}+1}}$.

As such, we have $\mathbb{E} \{\| \sum\limits_{s=1}^S \sqrt{\eta_{ks}} \|\mathbf{h}_{ks}\|^2 \| ^2 \}= \mathbb{E} \big{\{}\sum\limits_{s=1}^S \sum\limits_{t=1}^S \sqrt{\eta_{ks} \eta_{kt}} \big{\{} \|\bar{\mathbf{h}}_{ks}\|^2\|\bar{\mathbf{h}}_{kt}\|^2+ 2\|\bar{\mathbf{h}}_{ks}\|^2 \sum\limits_{n=1}^{4}\|\mathbf{x}_{n,kt}\|^2 \\ + 2\text{Re}(\bar{\mathbf{h}}_{ks}^H \mathbf{x}_{3,ks}  \mathbf{x}_{3,kt}^H \bar{\mathbf{h}}_{kt} +2\bar{\mathbf{h}}_{ks}^H\mathbf{x}_{3,ks} \mathbf{x}_{4,kt}^H \mathbf{x}_{2,kt} + \mathbf{x}_{2,ks}^H \mathbf{x}_{4,ks} \mathbf{x}_{4,kt}^H \mathbf{x}_{2,kt} )  + \sum\limits_{n_1=1}^{4} \sum\limits_{n_2=1}^{4} \|\mathbf{x}_{n_1,ks}\|^2
\|\mathbf{x}_{n_2,kt}\|^2 \big{\}}+\sum\limits_{s=1}^S \eta_{ks} \big{\{} \bar{\mathbf{h}}_{ks}^H (\sum\limits_{n=1,2,4} \mathbf{x}_{n,ks}\mathbf{x}_{n,ks}^H)\bar{\mathbf{h}}_{ks}  +  \sum_{(n_1,n_2)\neq\{(2,4),(4,2)\}, n_1 \neq n_2}^{4,4} |\mathbf{x}_{n_1,ks}^H
\cdot\mathbf{x}_{n_2,ks}|^2 \big{\}}$.
\subsubsection{Calculate $\mathbb{E} \{\sum\limits_{n=1}^{4}\|\mathbf{x}_{n,ks}\|^2\}$}
\begin{enumerate}[(i)]
	\item $\mathbb{E}\{\|\mathbf{x}_{\mathrm{1},ks}\|^2\}=b_{\mathrm{d},ks}^2\mathbb{E}\{\tilde{\mathbf{h}}_{\mathrm{d},k s}^T \tilde{\mathbf{h}}_{\mathrm{d},k s}^*\}=M b_{\mathrm{d},ks}^2$.
	\item
	$\mathbb{E}\{\|\mathbf{x}_{\mathrm{2},ks}\|^2\}=\sum\limits_{l=1}^{L}{a_{\mathrm{r},lk}^2b_{ls}^2 \bar{\mathbf{h}}_{\mathrm{r},l k}}^H \bm{\Theta}_{l}^{*} \mathbb{E}\{\tilde{\mathbf{G}}_{ls}^{*}\tilde{\mathbf{G}}_{ls}^{T}\} \bm{\Theta}_{l}^{T}\bar{\mathbf{h}}_{\mathrm{r},l k}=MN\sum\limits_{l=1}^{L}{a_{\mathrm{r},lk}^2b_{ls}^2}$.
	\item$\mathbb{E}\{\|\mathbf{x}_{\mathrm{3},ks}\|^2\}=\sum\limits_{l=1}^{L}\sum\limits_{m=1}^{L} b_{\mathrm{r},lk}a_{ls}b_{\mathrm{r},mk}a_{ms} \mathbb{E}\{\tilde{\mathbf{h}}_{\mathrm{r},l k}^H \bm{\Theta}_{l}^{*} \bar{\mathbf{G}}_{ls}^{*}\bar{\mathbf{G}}_{ms}^{T}\bm{\Theta}_{m}^{T}\tilde{\mathbf{h}}_{\mathrm{r},m k} \}$.
	
	Noticing that when $\mathbf{H} \in \mathbb{C}^{N_l \times N_t}$ is i.i.d complex Gaussian random matrix with zero mean and unit variance, the expectation can be calculated as $\mathbb{E} \{ \mathbf{H}^H \mathbf{A} \mathbf{H} \} = \text{Tr}(\mathbf{A}) \mathbf{I}_{N_t}$ for any matrix $\mathbf{A} \in \mathbb{C}^{N_l \times N_l}$. Hence, $\mathbb{E}\{\|\mathbf{x}_{\mathrm{3},ks}\|^2\}=\sum\limits_{l=1}^{L} b_{\mathrm{r},lk}^2 a_{ls}^2 \text{tr}(\bm{\Theta}_{l}^{*} \bar{\mathbf{G}}_{ls}^{*}\bar{\mathbf{G}}_{ms}^{T}\bm{\Theta}_{m}^{T})=MN\sum\limits_{l=1}^{L} b_{\mathrm{r},lk}^2 a_{ls}^2$.
	
	\item $\mathbb{E}\{\|\mathbf{x}_{\mathrm{4},ks}\|^2\}=\sum\limits_{l=1}^{L}b_{\mathrm{r},lk}^2b_{ls}^2\text{tr}(\bm{\Theta}_{l}^{*} \mathbb{E}\{ \tilde{\mathbf{G}}_{ls}^{*} \tilde{\mathbf{G}}_{ls}^{T} \}\bm{\Theta}_{l}^{T})=MN\sum\limits_{l=1}^{L}b_{\mathrm{r},lk}^2b_{ls}^2$.
\end{enumerate}
To conclude, $\sum\limits_{s=1}^S \sqrt{\eta_{ks}} \mathbb{E} \big{\{} \|\mathbf{h}_{ks}^T\|^2\big{\}}=\sum\limits_{s=1}^S \sqrt{\eta_{ks}}(\|\bar{\mathbf{h}}_{ks}\|^2+ M \chi_{ks})$, where $\chi_{ks}=b_{\mathrm{d},ks}^2+N(\alpha_{1,ks}+\alpha_{2,ks}+\alpha_{3,ks})$, $\alpha_{1,ks}=\sum\limits_{l=1}^{L}{a_{\mathrm{r},lk}^2b_{ls}^2}$, $\alpha_{2,ks}=\sum\limits_{l=1}^{L} b_{\mathrm{r},lk}^2 a_{ls}^2$ and $\alpha_{3,ks}=\sum\limits_{l=1}^{L}b_{\mathrm{r},lk}^2b_{ls}^2$.

\subsubsection{Calculate}
$\sum_{s=1}^S \sum_{t=1}^S \sqrt{\eta_{ks} \eta_{kt}}\mathbb{E} \{\mathbf{x}_{2,ks}^H \mathbf{x}_{4,ks} \mathbf{x}_{4,kt}^H \mathbf{x}_{2,kt}\}=\sum_{s=1}^S \sum_{t=1}^S \sqrt{\eta_{ks} \eta_{kt}}\sum_{l=1}^{L}\sum_{m=1}^{L} \\ \sum_{v=1}^{L}
a_{\mathrm{r},l k}b_{ls} b_{\mathrm{r},mk}^2b_{ms} b_{mt}a_{\mathrm{r},vk}b_{vt}\bar{\mathbf{h}}_{\mathrm{r},l k}^H \bm{\Theta}_{l}^{*} \mathbb{E} \big{\{} \tilde{\mathbf{G}}_{ls}^{*} \tilde{\mathbf{G}}_{ms}^{T} \tilde{\mathbf{G}}_{mt}^{*} \tilde{\mathbf{G}}_{vt}^{T}  \big{\}}\bm{\Theta}_{v}^{T}\bar{\mathbf{h}}_{\mathrm{r},v k}$.
\begin{enumerate}[(i)]
	\item When $s \neq t$, we have $\sum_{s=1}^S \sum_{t\neq s}^S \sqrt{\eta_{ks} \eta_{kt}}\sum_{l=1}^{L}a_{\mathrm{r},l k}^2b_{ls}^2 b_{\mathrm{r},lk}^2 b_{lt}^2\bar{\mathbf{h}}_{\mathrm{r},l k}^H \bm{\Theta}_{l}^{*} \mathbb{E} \big{\{} \tilde{\mathbf{G}}_{ls}^{*} \tilde{\mathbf{G}}_{ls}^{T} \tilde{\mathbf{G}}_{lt}^{*} \tilde{\mathbf{G}}_{lt}^{T}  \big{\}}\bm{\Theta}_{v}^{T}\\ \bar{\mathbf{h}}_{\mathrm{r},v k}=M^2 N\sum_{s=1}^S \sum_{t\neq s}^S \sqrt{\eta_{ks} \eta_{kt}}\sum_{l=1}^{L}a_{\mathrm{r},l k}^2b_{ls}^2 b_{\mathrm{r},lk}^2 b_{lt}^2$.
	
	\item When $s=t$, we have $\sum\limits_{s=1}^S  \eta_{ks} \sum\limits_{l=1}^{L}\sum\limits_{m=1}^{L}
	a_{\mathrm{r},l k}^2b_{ls}^2 b_{\mathrm{r},mk}^2b_{ms}^2\bar{\mathbf{h}}_{\mathrm{r},l k}^H \bm{\Theta}_{l}^{*} \mathbb{E} \big{\{} \tilde{\mathbf{G}}_{ls}^{*} \tilde{\mathbf{G}}_{ms}^{T} \tilde{\mathbf{G}}_{ms}^{*} \tilde{\mathbf{G}}_{ls}^{T}  \big{\}}\bm{\Theta}_{l}^{T}\bar{\mathbf{h}}_{\mathrm{r},l k}$.

	\begin{enumerate}[(a)]
		\item When $m \neq l$, we obtain $N^2 M\sum_{s=1}^S  \eta_{ks} \sum_{l=1}^{L}\sum_{m\neq l}^{L} a_{\mathrm{r},l k}^2b_{ls}^2 b_{\mathrm{r},mk}^2b_{ms}^2$.
		
		\item When $m = l$, let $\tilde{\mathbf{g}}_{i,ls}$ denote the $i$-th row of the $\tilde{\mathbf{G}}_{ls}$ and follow the i.i.d. complex Gaussian distribution, we have $\mathbb{E}\{\|\tilde{\mathbf{g}}_{i,ls}\|^2\}=M$ and $\mathbb{E}\{\|\tilde{\mathbf{g}}_{i,ls}\|^4\}=M^2+M$. Hence, we obtain $\mathbb{E} \big{\{} \tilde{\mathbf{G}}_{ls}^{*} \tilde{\mathbf{G}}_{ls}^{T} \tilde{\mathbf{G}}_{ls}^{*} \tilde{\mathbf{G}}_{ls}^{T}\big{\}}=\mathbb{E} \big{\{} \text{diag}(\|\tilde{\mathbf{g}}_{1,ls}\|^2\sum_{j=1}^M\|\tilde{\mathbf{g}}_{j,ls}\|^2,...,\|\tilde{\mathbf{g}}_{N,ls}\|^2\sum_{j=1}^M\|\tilde{\mathbf{g}}_{j,ls}\|^2)  \big{\}}$. Thus, we calculate to get $MN(M+N)\sum_{s=1}^S  \eta_{ks} \sum_{l=1}^{L}a_{\mathrm{r},l k}^2b_{ls}^4 b_{\mathrm{r},lk}^2$.
	\end{enumerate}
\end{enumerate}

\subsubsection{Calculate}
$\sum_{s=1}^S \sum_{t\neq s}^S \sqrt{\eta_{ks} \eta_{kt}}\mathbb{E} \{\|\mathbf{x}_{3,ks}^H\|^2 \|\mathbf{x}_{3,kt}\|^2\}=\sum_{s=1}^S \sum_{t\neq s}^S \sqrt{\eta_{ks} \eta_{kt}} \big{[}\sum_{l=1}^{L} \sum_{p\neq l}^{L} b_{\mathrm{r},l k}^2 b_{\mathrm{r},p k}^2\\ a_{ls}^2 a_{pt}^2 \mathbb{E}\{\tilde{\mathbf{h}}_{\mathrm{r},l k}^H \bm{\Theta}_{l}^{*} \bar{\mathbf{G}}_{ls}^{*}  \bar{\mathbf{G}}_{ls}^{T}\bm{\Theta}_{l}^{T} \tilde{\mathbf{h}}_{\mathrm{r},l k} \tilde{\mathbf{h}}_{\mathrm{r},p k}^H \bm{\Theta}_{p}^{*} \bar{\mathbf{G}}_{pt}^{*} \bar{\mathbf{G}}_{pt}^{T}\bm{\Theta}_{p}^{T}\tilde{\mathbf{h}}_{\mathrm{r},p k}\}+\sum_{l=1}^{L} \sum_{m\neq l}^{L}  b_{\mathrm{r},l k}^2 b_{\mathrm{r},m k}^2a_{ls}a_{lt}a_{ms}a_{mt}\mathbb{E}\{\tilde{\mathbf{h}}_{\mathrm{r},l k}^H \bm{\Theta}_{l}^{*} \\ \bar{\mathbf{G}}_{ls}^{*} \bar{\mathbf{G}}_{ms}^{T}\bm{\Theta}_{m}^{T} \tilde{\mathbf{h}}_{\mathrm{r},m k} \tilde{\mathbf{h}}_{\mathrm{r},m k}^H \bm{\Theta}_{m}^{*} \bar{\mathbf{G}}_{mt}^{*} \bar{\mathbf{G}}_{lt}^{T}\bm{\Theta}_{l}^{T}\tilde{\mathbf{h}}_{\mathrm{r},l k}\}+\sum_{l=1}^{L} b_{\mathrm{r},l k}^4 a_{ls}^2 a_{lt}^2 \mathbb{E}\{\tilde{\mathbf{h}}_{\mathrm{r},l k}^H \bm{\Theta}_{l}^{*} \bar{\mathbf{G}}_{ls}^{*} \bar{\mathbf{G}}_{ls}^{T}\bm{\Theta}_{l}^{T} \tilde{\mathbf{h}}_{\mathrm{r},l k} \tilde{\mathbf{h}}_{\mathrm{r},l k}^H \bm{\Theta}_{l}^{*} \bar{\mathbf{G}}_{lt}^{*} \bar{\mathbf{G}}_{lt}^{T}\\ \bm{\Theta}_{l}^{T}\tilde{\mathbf{h}}_{\mathrm{r},l k}\} \big{]}=\sum_{l=1}^{L}  \sum_{m=1}^{L}  b_{\mathrm{r},l k}^2 b_{\mathrm{r},m k}^2 a_{ls}a_{lt}a_{ms}a_{mt} \text{tr}(\bar{\mathbf{G}}_{ls}^{*} \bar{\mathbf{G}}_{ms}^{T}\bar{\mathbf{G}}_{mt}^{*}\bar{\mathbf{G}}_{lt}^{T})+M^2N^2\alpha_{2,ks} \alpha_{2,kt}$.

\subsubsection{Calculate}
$\sum_{s=1}^S \sum_{t\neq s}^S \sqrt{\eta_{ks} \eta_{kt}} \mathbb{E}\{\|\mathbf{x}_{4,ks}^H\|^2\|\mathbf{x}_{4,kt}\|^2\}=\sum_{s=1}^S \sum_{t\neq s}^S \sqrt{\eta_{ks} \eta_{kt}} \big{[}\sum_{l=1}^{L} \sum_{m \neq l}^{L} b_{\mathrm{r},lk}^2 b_{ls}^2\\ b_{\mathrm{r},mk}^2 b_{mt}^2 \mathbb{E}\{ \tilde{\mathbf{h}}_{\mathrm{r},l k}^H \bm{\Theta}_{l}^{*}  \tilde{\mathbf{G}}_{ls}^{*}\tilde{\mathbf{G}}_{ls}^{T}\bm{\Theta}_{l}^{T}\tilde{\mathbf{h}}_{\mathrm{r},l k} \tilde{\mathbf{h}}_{\mathrm{r},mk}^H \bm{\Theta}_{m}^{*}\tilde{\mathbf{G}}_{mt}^{*} \tilde{\mathbf{G}}_{mt}^{T}\bm{\Theta}_{m}^{T}\tilde{\mathbf{h}}_{\mathrm{r},m k}  + \sum_{l=1}^{L} b_{\mathrm{r},lk}^4 b_{ls}^2 b_{lt}^2 \mathbb{E}\{ \tilde{\mathbf{h}}_{\mathrm{r},l k}^H  \bm{\Theta}_{l}^{*}\tilde{\mathbf{G}}_{ls}^{*} \tilde{\mathbf{G}}_{ls}^{T}\bm{\Theta}_{l}^{T}\tilde{\mathbf{h}}_{\mathrm{r},l k}\\ \tilde{\mathbf{h}}_{\mathrm{r},lk}^H  \bm{\Theta}_{l}^{*}\tilde{\mathbf{G}}_{lt}^{*}\tilde{\mathbf{G}}_{lt}^{T} \bm{\Theta}_{l}^{T}\tilde{\mathbf{h}}_{\mathrm{r},l k}\} \big{]}=M^2 N^2 \alpha_{3,ks} \alpha_{3,kt}+M^2 N\sum_{l=1}^{L} b_{\mathrm{r},lk}^4 b_{ls}^2 b_{lt}^2$.

\subsubsection{Similarly, we have} $\sum_{s=1}^S  \eta_{ks}\mathbb{E}\{ \sum_{n_1=1}^{4} \|\mathbf{x}_{n_1,ks}^H\|^4\} =\sum_{s=1}^S  \eta_{ks} \big{[} (M^2+M)(b_{\mathrm{d},ks}^4+N^2\alpha_{1,ks}^2+N^2\alpha_{2,ks}^2+N^2\alpha_{3,ks}^2)-N^2M\alpha_{2,ks}^2+\sum_{l=1}^{L}\sum_{m=1}^{L}a_{ls}^2a_{ms}^2b_{\mathrm{r},lk}^2b_{\mathrm{r},mk}^2\|\bar{\mathbf{G}}_{ls}^T\bar{\mathbf{G}}_{ms}^*\|^2+N(M^2+M)\sum_{m=1}^{L}b_{ls}^4b_{\mathrm{r},lk}^4 \big{]}$.

\subsubsection{Calculate}
$\sum_{s=1}^S  \eta_{ks} \mathbb{E}\{|\mathbf{x}_{2,ks}^H \mathbf{x}_{4,ks}|^2\}=\sum_{s=1}^S  \eta_{ks} \mathbb{E}\{ \sum_{l=1}^{L} \sum_{m\neq l}^{L}a_{\mathrm{r},l k}^2b_{ls}^2 b_{\mathrm{r},m k}^2b_{ms}^2 \bar{\mathbf{h}}_{\mathrm{r},l k}^H \bm{\Theta}_{l}^{*} \tilde{\mathbf{G}}_{ls}^{*} \tilde{\mathbf{G}}_{ms}^{T}\\ \bm{\Theta}_{m}^{T}\tilde{\mathbf{h}}_{\mathrm{r},mk} \tilde{\mathbf{h}}_{\mathrm{r},m k}^H \bm{\Theta}_{m}^{*} \tilde{\mathbf{G}}_{ms}^{*} \tilde{\mathbf{G}}_{ls}^{T}\bm{\Theta}_{l}^{T}\bar{\mathbf{h}}_{\mathrm{r},l k} + \sum_{l=1}^{L} a_{\mathrm{r},l k}^2b_{ls}^4 b_{\mathrm{r},m k}^2 \bar{\mathbf{h}}_{\mathrm{r},l k}^H \bm{\Theta}_{l}^{*} \tilde{\mathbf{G}}_{ls}^{*} \tilde{\mathbf{G}}_{ls}^{T}\bm{\Theta}_{l}^{T}\tilde{\mathbf{h}}_{\mathrm{r},lk} \tilde{\mathbf{h}}_{\mathrm{r},lk}^H \bm{\Theta}_{l}^{*} \tilde{\mathbf{G}}_{ls}^{*} \tilde{\mathbf{G}}_{ls}^{T}\bm{\Theta}_{l}^{T}\bar{\mathbf{h}}_{\mathrm{r},l k}\}= \sum_{s=1}^S  \eta_{ks}(N^2M\alpha_{1,ks}\alpha_{3,ks}+M^2N\sum_{l=1}^{L} a_{\mathrm{r},l k}^2b_{\mathrm{r},l k}^2b_{ls}^4)$.

\subsubsection{Calculate}
$\sum_{s=1}^S  \eta_{ks}\mathbb{E}\{|\mathbf{x}_{3,ks}^H\mathbf{x}_{4,ks}|^2\}=\sum_{s=1}^S  \eta_{ks}\mathbb{E}\{\sum_{l=1}^{L} \sum_{m\neq l}^{L} b_{\mathrm{r},l k}^2a_{ls}^2 b_{\mathrm{r},mk}^2b_{ms}^2\tilde{\mathbf{h}}_{\mathrm{r},l k}^H \bm{\Theta}_{l}^{*} \bar{\mathbf{G}}_{ls}^{*} \tilde{\mathbf{G}}_{ms}^{T}\\ \bm{\Theta}_{m}^{T}\tilde{\mathbf{h}}_{\mathrm{r},m k}\tilde{\mathbf{h}}_{\mathrm{r},m k}^H\bm{\Theta}_{m}^{*} \tilde{\mathbf{G}}_{ms}^{*} \bar{\mathbf{G}}_{ls}^{T}\bm{\Theta}_{l}^{T}\tilde{\mathbf{h}}_{\mathrm{r},l k} + \sum_{l=1}^{L} b_{\mathrm{r},l k}^4a_{ls}^2 b_{ls}^2\tilde{\mathbf{h}}_{\mathrm{r},l k}^H \bm{\Theta}_{l}^{*} \bar{\mathbf{G}}_{ls}^{*} \tilde{\mathbf{G}}_{ls}^{T}\bm{\Theta}_{l}^{T}\tilde{\mathbf{h}}_{\mathrm{r},lk}\tilde{\mathbf{h}}_{\mathrm{r},l k}^H\bm{\Theta}_{m}^{*} \tilde{\mathbf{G}}_{ls}^{*} \bar{\mathbf{G}}_{ls}^{T}\bm{\Theta}_{l}^{T}\tilde{\mathbf{h}}_{\mathrm{r},l k}\} =\sum_{s=1}^S  \eta_{ks} (N^2M\alpha_{2,ks} \alpha_{3,ks}+MN\sum_{l=1}^{L} b_{\mathrm{r},l k}^4a_{ls}^2 b_{ls}^2)$.

Other parts can be calculated in similar way, and omitted due to page limitation.

\subsection{Calculate $\mathbb{E} \{ \sum_{i \neq k}^K \|\sum_{s=1}^S \sqrt{\eta_{is}} \mathbf{h}_{ks}^T \mathbf{h}_{is}^{*} \|^2 \}$}
As such, it can be decomposed as follows:
$\sum_{s=1}^S \sum_{t=1}^S \sum_{i\neq k}^K \sqrt{\eta_{is} \eta_{it}} \mathbb{E} \big{\{}\bar{\mathbf{h}}_{ks}^T \bar{\mathbf{h}}_{is}^* \bar{\mathbf{h}}_{it}^T \bar{\mathbf{h}}_{kt}^*+\bar{\mathbf{h}}_{ks}^T \bar{\mathbf{h}}_{is}^* \mathbf{x}_{2,it}^T\\ \mathbf{x}_{2,kt}^* + \mathbf{x}_{2,ks}^T \mathbf{x}_{2,is}^* \bar{\mathbf{h}}_{it}^T \bar{\mathbf{h}}_{kt}^* +  \bar{\mathbf{h}}_{ks}^T \mathbf{x}_{3,is}^* \mathbf{x}_{3,it}^T \bar{\mathbf{h}}_{kt}^* + \mathbf{x}_{3,ks}^T  \bar{\mathbf{h}}_{is}^* \bar{\mathbf{h}}_{it}^T \mathbf{x}_{3,kt}^* + \bar{\mathbf{h}}_{ks}^T \mathbf{x}_{3,is}^* \mathbf{x}_{4,it}^T \mathbf{x}_{2,kt}^* + \mathbf{x}_{3,ks}^T \bar{\mathbf{h}}_{is}^* \mathbf{x}_{2,it}^T \mathbf{x}_{4,kt}^* +\mathbf{x}_{4,ks}^T \mathbf{x}_{2,is}^* \bar{\mathbf{h}}_{it}^T \mathbf{x}_{3,kt}^*+\mathbf{x}_{2,ks}^T \mathbf{x}_{4,is}^* \mathbf{x}_{3,it}^T  \bar{\mathbf{h}}_{kt}^* + \mathbf{x}_{3,ks}^T \mathbf{x}_{3,is}^* \mathbf{x}_{3,it}^T \mathbf{x}_{3,kt}^* + \mathbf{x}_{4,ks}^T \mathbf{x}_{4,is}^* \mathbf{x}_{3,it}^T \mathbf{x}_{3,kt}^* +\mathbf{x}_{3,ks}^T \mathbf{x}_{3,is}^* \mathbf{x}_{4,it}^T \mathbf{x}_{4,kt}^*+ \mathbf{x}_{2,ks}^T \mathbf{x}_{2,is}^* \mathbf{x}_{2,it}^T \mathbf{x}_{2,kt}^*+\mathbf{x}_{4,ks}^T \mathbf{x}_{4,is}^* \mathbf{x}_{4,it}^T \mathbf{x}_{4,kt}^*+\mathbf{x}_{2,ks}^T \mathbf{x}_{4,is}^* \mathbf{x}_{4,it}^T \mathbf{x}_{2,kt}^* +\mathbf{x}_{4,ks}^T \mathbf{x}_{2,is}^* \mathbf{x}_{2,it}^T \mathbf{x}_{4,kt}^*\big{\}}+ \sum_{s=1}^S \sum_{i\neq k}^K \eta_{is}\mathbb{E} \big{\{}\bar{\mathbf{h}}_{ks}^T\\ (\mathbf{x}_{1,is}^* \mathbf{x}_{1,is}^T+\mathbf{x}_{2,is}^* \mathbf{x}_{2,is}^T+\mathbf{x}_{4,is}^* \mathbf{x}_{4,is}^T) \bar{\mathbf{h}}_{is}^* + |\mathbf{x}_{1,ks}^T \mathbf{x}_{1,is}^*|^2 +|\mathbf{x}_{1,ks}^T \mathbf{x}_{2,is}^*|^2 +|\mathbf{x}_{2,ks}^T \mathbf{x}_{1,is}^*|^2 +|\mathbf{x}_{3,ks}^T \mathbf{x}_{1,is}^*|^2 +|\mathbf{x}_{3,ks}^T \mathbf{x}_{2,is}^*|^2 +|\mathbf{x}_{1,ks}^T \mathbf{x}_{3,is}^*|^2 +|\mathbf{x}_{2,ks}^T \mathbf{x}_{3,is}^*|^2 +|\mathbf{x}_{4,ks}^T \mathbf{x}_{1,is}^*|^2+|\mathbf{x}_{1,ks}^T \mathbf{x}_{4,is}^*|^2 +|\mathbf{x}_{3,ks}^T \mathbf{x}_{4,is}^*|^2  +|\mathbf{x}_{4,ks}^T \mathbf{x}_{3,is}^*|^2   \big{\}}$.

\subsubsection{Calculate}
$\sum\limits_{s=1}^S \sum\limits_{t=1}^S \sum\limits_{i\neq k}^K \sqrt{\eta_{is} \eta_{it}} \mathbb{E} \{  \mathbf{x}_{2,ks}^T \mathbf{x}_{2,is}^* \mathbf{x}_{2,it}^T \mathbf{x}_{2,kt}^* \} = \sum\limits_{s=1}^S \sum\limits_{t\neq s}^S \sum\limits_{i\neq k}^K \sqrt{\eta_{is} \eta_{it}}  \sum\limits_{l=1}^{L} \sum\limits_{m=1}^{L} a_{\mathrm{r},lk}a_{\mathrm{r},li} a_{\mathrm{r},mi} \\a_{\mathrm{r},mk}b_{ls}^2 b_{mt}^2 \bar{\mathbf{h}}_{\mathrm{r},l k}^T \bm{\Theta}_{l}  M \mathbf{I}_N \bm{\Theta}_{l}^{H} \bar{\mathbf{h}}_{\mathrm{r},l i}^* \bar{\mathbf{h}}_{\mathrm{r},m i}^T \bm{\Theta}_{m} M \mathbf{I}_N  \bm{\Theta}_{m}^H \bar{\mathbf{h}}_{\mathrm{r},m k}^*+ \sum\limits_{s=1}^S  \sum\limits_{i\neq k}^K \eta_{is}  \sum\limits_{l=1}^{L} \sum\limits_{m\neq l}^{L} a_{\mathrm{r},lk}a_{\mathrm{r},li}a_{\mathrm{r},mi}a_{\mathrm{r},mk}b_{ls}^2 b_{ms}^2 \bar{\mathbf{h}}_{\mathrm{r},l k}^T \\ \bar{\mathbf{h}}_{\mathrm{r},l i}^* \bar{\mathbf{h}}_{\mathrm{r},mi}^T \bar{\mathbf{h}}_{\mathrm{r},m k}^*+\sum\limits_{s=1}^S  \sum\limits_{i\neq k}^K \eta_{is}  \sum\limits_{l=1}^{L} \sum\limits_{m\neq l}^{L} a_{\mathrm{r},lk}^2 a_{\mathrm{r},mi}^2 b_{ls}^2 b_{ms}^2 \bar{\mathbf{h}}_{\mathrm{r},l k}^T \bm{\Theta}_{l} \text{tr} \big{(}\text{tr}(\bm{\Theta}_{m}^H \bar{\mathbf{h}}_{\mathrm{r},mi}^* \bar{\mathbf{h}}_{\mathrm{r},mi}^T \bm{\Theta}_{m})\mathbf{I}_M  \big{)}\mathbf{I}_N  \bm{\Theta}_{l}^H \bar{\mathbf{h}}_{\mathrm{r},lk}^*+\sum\limits_{s=1}^S  \sum\limits_{i\neq k}^K \\ \eta_{is}  \sum\limits_{l=1}^{L} a_{\mathrm{r},lk}^2  a_{\mathrm{r},mi}^2 b_{ls}^4 \bar{\mathbf{h}}_{\mathrm{r},lk}^T \bm{\Theta}_{l} \mathbb{E}\{ \tilde{\mathbf{G}}_{ls} \tilde{\mathbf{G}}_{ls}^{H}\bm{\Theta}_{l}^H \bar{\mathbf{h}}_{\mathrm{r},li}^* \bar{\mathbf{h}}_{\mathrm{r},li}^T \bm{\Theta}_{l}\tilde{\mathbf{G}}_{ls} \tilde{\mathbf{G}}_{ls}^{H} \}\bm{\Theta}_{l}^H \bar{\mathbf{h}}_{\mathrm{r},lk}^* =  M^2 \sum\limits_{s=1}^S \sum\limits_{t=1}^S \sum\limits_{i\neq k}^K \sqrt{\eta_{is} \eta_{it}}  \sum\limits_{l=1}^{L} \sum\limits_{m=1}^{L} a_{\mathrm{r},lk}\\ a_{\mathrm{r},li}a_{\mathrm{r},mi}a_{\mathrm{r},mk}b_{ls}^2 b_{mt}^2  \bar{\mathbf{h}}_{\mathrm{r},l k}^T \bar{\mathbf{h}}_{\mathrm{r},l i}^* \bar{\mathbf{h}}_{\mathrm{r},mi}^T \bar{\mathbf{h}}_{\mathrm{r},m k}^*+ \sum\limits_{s=1}^S  \sum\limits_{i\neq k}^K \eta_{is} N^2M\alpha_{1,ks}\alpha_{1,is}$.

\subsubsection{Calculate}
$\sum_{s=1}^S \sum_{t=1}^S \sum_{i\neq k}^K \sqrt{\eta_{is} \eta_{it}} \mathbb{E} \{  \mathbf{x}_{4,ks}^T \mathbf{x}_{4,is}^* \mathbf{x}_{4,it}^T \mathbf{x}_{4,kt}^*+\mathbf{x}_{4,ks}^T \mathbf{x}_{2,is}^* \mathbf{x}_{2,it}^T \mathbf{x}_{4,kt}^* + \mathbf{x}_{2,ks}^T \mathbf{x}_{4,is}^* \mathbf{x}_{4,it}^T \mathbf{x}_{2,kt}^* \}\\ = M^2N\sum_{s=1}^S \sum_{t=1}^S \sum_{i\neq k}^K \sqrt{\eta_{is} \eta_{it}} \sum_{l=1}^{L}b_{ls}^2 b_{lt}^2 (b_{\mathrm{r},lk}^2b_{\mathrm{r},li}^2+a_{\mathrm{r},li}^2b_{\mathrm{r},lk}^2+a_{\mathrm{r},lk}^2 b_{\mathrm{r},li}^2) + N^2 M \sum_{s=1}^S \sum_{i\neq k}^K (\alpha_{3,ks}\alpha_{3,is}+\alpha_{3,ks}\alpha_{1,is}+\alpha_{1,ks}\alpha_{3,is})$.

\subsubsection{Similarly, we have}
$\sum_{s=1}^S \sum_{i\neq k}^K \eta_{is}\mathbb{E} \big{\{}\bar{\mathbf{h}}_{ks}^T (\mathbf{x}_{1,is}^* \mathbf{x}_{1,is}^T+\mathbf{x}_{2,is}^* \mathbf{x}_{2,is}^T+\mathbf{x}_{4,is}^* \mathbf{x}_{4,is}^T) \bar{\mathbf{h}}_{is}^* + |\mathbf{x}_{1,ks}^T \mathbf{x}_{1,is}^*|^2 +|\mathbf{x}_{1,ks}^T \mathbf{x}_{2,is}^*|^2 +|\mathbf{x}_{2,ks}^T \mathbf{x}_{1,is}^*|^2 +|\mathbf{x}_{3,ks}^T \mathbf{x}_{1,is}^*|^2 +|\mathbf{x}_{3,ks}^T \mathbf{x}_{2,is}^*|^2 +|\mathbf{x}_{1,ks}^T \mathbf{x}_{3,is}^*|^2 +|\mathbf{x}_{2,ks}^T \mathbf{x}_{3,is}^*|^2 +|\mathbf{x}_{4,ks}^T \mathbf{x}_{1,is}^*|^2+|\mathbf{x}_{1,ks}^T \mathbf{x}_{4,is}^*|^2 +|\mathbf{x}_{3,ks}^T \mathbf{x}_{4,is}^*|^2  +|\mathbf{x}_{4,ks}^T \mathbf{x}_{3,is}^*|^2   \big{\}} = \sum_{s=1}^S \sum_{i\neq k}^K \eta_{is} \big{[} \|\bar{\mathbf{h}}_{ks}\|^2(b_{\mathrm{d},is}^2+N\alpha_{1,is}+N\alpha_{3,is})+\|\bar{\mathbf{h}}_{is}\|^2(b_{\mathrm{d},ks}^2+N\alpha_{1,ks}+N\alpha_{3,ks})+Mb_{\mathrm{d},ks}^2b_{\mathrm{d},is}^2+MNb_{\mathrm{d},ks}^2(\alpha_{1,is}+\alpha_{2,is}+\alpha_{3,is})+MNb_{\mathrm{d},is}^2(\alpha_{1,ks}+\alpha_{2,ks}+\alpha_{3,ks})+N^2M(\alpha_{1,ks}\alpha_{2,is}+\alpha_{2,ks}\alpha_{1,is}+\alpha_{2,ks}\alpha_{3,is}+\alpha_{3,ks}\alpha_{2,is})\big{]}$.

Other parts can be calculated in similar way, and omitted due to page limitation.

\section{The Equivalent Transform to MSE Minimization Problem}\label{MMSE_eq}
Invoking the MMSE receiver $r_k$ in the Eq. (\ref{MSE}), we obtain
\begin{equation}
\mathbb{E}\{ \text{error}_k \} =  \frac{\sum\limits_{i \neq k}^K \mathcal{B}_{k,i} + N_0}{\mathcal{A}_k+\sum\limits_{i \neq k}^K \mathcal{B}_{k,i} + N_0}.
\end{equation}

Then, the following useful relation holds:
\begin{equation}
\bar{C}_k = \log_2(\mathbb{E}\{ \text{error}_k \}^{-1}).
\end{equation}

After introducing auxiliary variable $\kappa_k$ and applying the Lagrangian dual transform proposed in \cite{fractional}, the expression for $\bar{C}_k$ can be equivalently written as
\begin{equation}
	\bar{C}_k = \log_2(\kappa_k)- \kappa_k \mathbb{E}\{\text{error}_k\} + 1,
\end{equation}
where $\kappa_k$ can be obtained by setting $\partial \bar{C}_k/ \partial \kappa_k$ to zero. Then, fixing other variables, the maximization problem of $\bar{C}_k$ can be reformulated to the minimization problem of $\kappa_k \mathbb{E}\{\text{error}_k\}$.

\section{The Transformation Process of the PDD-based Method on the optimization $(P4)$}\label{PDD}
In order to transform it to a convex function, we first introduce a auxiliary vector: $\mathbf{x}$ with equility constraints: $\mathbf{x}=\mathbf{u}$, which aims to render the variables in the reformulated problem jointly convex. Specifically, several $\{\mathbf{u}\}$ are replaced symmetrically with $\{\mathbf{x}\}$, which makes the objective function to be quadratic program w.r.t. $\mathbf{u}$ and $\mathbf{x}$, respectively. Therefore, $\mathcal{F}(\mathbf{u},\mathbf{x})$ is guaranteed to be equivalent to objective function of $(P3)$ by discarding irrelevant variables with the equility constraints, where
\begin{align}
&\mathcal{F}(\mathbf{u}, \mathbf{x})=\textstyle\sum\limits\limits_{k=1}^K \kappa_k \mu_k |r_k|^2 \bigg{\{} \sum\limits\limits_{s=1}^S \sum\limits_{t=1}^S \sum\limits_{i=1}^K \sqrt{\eta_{is} \eta_{it}} \big{[} (\overline{\mathbf{GG}}_s \overline{\mathbf{hh}}_{r,k} \mathbf{u} + \overline{\mathbf{hh}}_{d,ks})^T (\overline{\mathbf{GG}}_s\overline{\mathbf{hh}}_{r,i} \mathbf{x} + \overline{\mathbf{hh}}_{d,is})^*
(\overline{\mathbf{GG}}_t \notag\\
&\overline{\mathbf{hh}}_{r,i} \mathbf{x} + \overline{\mathbf{hh}}_{d,it})^T (\overline{\mathbf{GG}}_t \overline{\mathbf{hh}}_{r,k} \mathbf{u} + \overline{\mathbf{hh}}_{d,kt})^* + (\overline{\mathbf{GG}}_s \overline{\mathbf{hh}}_{r,k} \mathbf{u} + \overline{\mathbf{hh}}_{d,ks})^T \overline{\mathbf{GG}}_{is}^* \overline{\mathbf{GG}}_{it}^T  (\overline{\mathbf{GG}}_t
\overline{\mathbf{hh}}_{r,k} \mathbf{u} +\notag\\
&\overline{\mathbf{hh}}_{d,kt})^* + (\overline{\mathbf{GG}}_t \overline{\mathbf{hh}}_{r,i} \mathbf{u} + \overline{\mathbf{hh}}_{d,it})^T \overline{\mathbf{GG}}_{kt}^* \overline{\mathbf{GG}}_{ks}^T (\overline{\mathbf{GG}}_s \overline{\mathbf{hh}}_{r,i} \mathbf{u} + \overline{\mathbf{hh}}_{d,is})^*+2M \text{Re}\big{(}
(\overline{\mathbf{GG}}_s \overline{\mathbf{hh}}_{r,k} \mathbf{u} + \notag\\
& \overline{\mathbf{hh}}_{d,ks})^T (\overline{\mathbf{GG}}_s \overline{\mathbf{hh}}_{r,i} \mathbf{x} + \overline{\mathbf{hh}}_{d,is})^* \widehat{\mathbf{hh}}_{r,it} \widehat{\mathbf{hh}}_{r,kt}^H + (\overline{\mathbf{GG}}_s \overline{\mathbf{hh}}_{r,k} \mathbf{u} + \overline{\mathbf{hh}}_{d,ks})^T \widehat{\mathbf{GG}}_{is}^*
\overline{\mathbf{hh}}_{r,kt}^* \mathbf{x}^* + (\overline{\mathbf{GG}}_s \overline{\mathbf{hh}}_{r,i}\notag\\
& \mathbf{u} + \overline{\mathbf{hh}}_{d,is})^T \widehat{\mathbf{GG}}_{ks}^* \overline{\mathbf{hh}}_{r,it}^* \mathbf{x}^* \big{)}\big{]} +\textstyle\sum\limits_{s=1}^S \sum\limits_{t=1}^S \sqrt{\eta_{ks}\eta_{kt}} \big{[} 2M(b_{d,kt}^2+N\alpha_{2,kt}
+N\alpha_{3,kt})(\overline{\mathbf{GG}}_s  \overline{\mathbf{hh}}_{r,k} \mathbf{u} + \overline{\mathbf{hh}}_{d,ks})^T \notag\\
&(\overline{\mathbf{GG}}_s \overline{\mathbf{hh}}_{r,k} \mathbf{u} + \overline{\mathbf{hh}}_{d,ks})^* \big{]} + \textstyle\sum\limits_{s=1}^S \sum\limits_{i=1}^K \eta_{is} \big{[} (b_{d,is}^2+N \alpha_{1,is}
+N \alpha_{3,is})(\overline{\mathbf{GG}}_s \overline{\mathbf{hh}}_{r,k} \mathbf{u} + \overline{\mathbf{hh}}_{d,ks})^T (\overline{\mathbf{GG}}_s \overline{\mathbf{hh}}_{r,k}\notag\\
& \mathbf{u} + \overline{\mathbf{hh}}_{d,ks})^* + (b_{d,ks}^2+N \alpha_{1,ks}+N \alpha_{3,ks})
(\overline{\mathbf{GG}}_s \overline{\mathbf{hh}}_{r,i} \mathbf{u} + \overline{\mathbf{hh}}_{d,is})^T (\overline{\mathbf{GG}}_s \overline{\mathbf{hh}}_{r,i} \mathbf{u} + \overline{\mathbf{hh}}_{d,is})^*\big{]} - \textstyle\sum\limits_{s=1}^S \eta_{ks}\notag\\
&\big{[}(b_{d,ks}^2+N \alpha_{1,ks}+N \alpha_{3,ks}) \text{Re} \big{(}
(\overline{\mathbf{GG}}_s \overline{\mathbf{hh}}_{r,k} \mathbf{u} + \overline{\mathbf{hh}}_{d,ks})^T (\overline{\mathbf{GG}}_s \overline{\mathbf{hh}}_{r,k} \mathbf{x} + \overline{\mathbf{hh}}_{d,ks})^* \big{)}\big{]}\bigg{\}} - 2 \textstyle\sum\limits_{k=1}^K \kappa_k \mu_k \notag\\
&\text{Re} \big{(} r_k \textstyle\sum\limits_{s=1}^S \sqrt{\eta_{ks}} (\overline{\mathbf{GG}}_s \overline{\mathbf{hh}}_{r,k} \mathbf{u}
+ \overline{\mathbf{hh}}_{d,ks})^T (\overline{\mathbf{GG}}_s \overline{\mathbf{hh}}_{r,k} \mathbf{x} + \overline{\mathbf{hh}}_{d,ks})^* \big{)},
\end{align}
$\overline{\mathbf{GG}}_s=[a_{1s}\bar{\mathbf{G}}_{1s}^T,...,a_{Ls}\bar{\mathbf{G}}_{Ls}^T]$, $\overline{\mathbf{GG}}_{ks}=[a_{1s}b_{r,1k}\bar{\mathbf{G}}_{1s}^T,...,a_{Ls}b_{r,Lk}\bar{\mathbf{G}}_{Ls}^T]$, $\widehat{\mathbf{GG}}_{ks}=[a_{1s}b_{r,1k}^2\bar{\mathbf{G}}_{1s}^T,...,a_{Ls}\\ b_{r,Lk}^2 \bar{\mathbf{G}}_{Ls}^T]$, $\overline{\mathbf{hh}}_{r,k}=\text{diag}\big{(} a_{r,1k}\text{diag}(\bar{\mathbf{h}}_{r,1k}),...,a_{r,Lk}\text{diag}(\bar{\mathbf{h}}_{r,Lk})\big{)}$, $\overline{\mathbf{hh}}_{r,ks}=\text{diag}\big{(}a_{r,1k}b_{1s}^2\text{diag}(\bar{\mathbf{h}}_{r,1k}),...,\\ a_{r,Lk} b_{Ls}^2\text{diag}(\bar{\mathbf{h}}_{r,Lk})\big{)}$, $\widehat{\mathbf{hh}}_{r,ks}=[b_{1s}a_{r,1k}\bar{\mathbf{h}}_{r,1k}^T,...,b_{Ls}a_{r,Lk}\bar{\mathbf{h}}_{r,Lk}^T]$ and $\overline{\mathbf{hh}}_{d,ks} = a_{d,ks} \bar{\mathbf{h}}_{d,ks}$.

On the other hand, we introduce $\mathbf{z}$ with equility constraints: $\mathbf{z}=\mathbf{u}$ to handle the unit modulus constraint. Finally, we will transform the optimization $(P3)$ to $(P4)$ based on the PDD method.

\section{The Definition of $\mathbf{R}$, $\mathbf{t}$, $\mathbf{Q}$ and $\mathbf{a}$}\label{definition} \vspace{-30pt}
\begin{align}
&\mathbf{R}=\textstyle\sum\limits_{k=1}^K \kappa_k \mu_k |r_k|^2 \bigg{\{} \sum\limits_{s=1}^S \sum\limits_{t=1}^S \sum\limits_{i=1}^K \sqrt{\eta_{is} \eta_{it}} \big{[} \overline{\mathbf{hh}}_{r,k}^T\overline{\mathbf{GG}}_s^T (\overline{\mathbf{GG}}_s \overline{\mathbf{hh}}_{r,i} \mathbf{x} + \overline{\mathbf{hh}}_{d,is})^* (\overline{\mathbf{GG}}_t \overline{\mathbf{hh}}_{r,i} \mathbf{x} + \overline{\mathbf{hh}}_{d,it})^T \notag\\
& \overline{\mathbf{GG}}_t^* \overline{\mathbf{hh}}_{r,k}^* +\overline{\mathbf{hh}}_{r,k}^T \overline{\mathbf{GG}}_s^T \overline{\mathbf{GG}}_{is}^* \overline{\mathbf{GG}}_{it}^T \overline{\mathbf{GG}}_{t}^* \overline{\mathbf{hh}}_{r,k}^* +\overline{\mathbf{hh}}_{r,i}^T \overline{\mathbf{GG}}_t^T \overline{\mathbf{GG}}_{kt}^* \overline{\mathbf{GG}}_{ks}^T \overline{\mathbf{GG}}_{s}^* \overline{\mathbf{hh}}_{r,i}^* \big{]}+ 2M \textstyle\sum\limits_{s=1}^S \sum\limits_{t=1}^S \notag\\
&\sqrt{\eta_{ks} \eta_{kt}} (b_{d,kt}^2+N\alpha_{2,kt}+N\alpha_{3,kt})\overline{\mathbf{hh}}_{r,k}^T \overline{\mathbf{GG}_s}^T \overline{\mathbf{GG}}_s^* \overline{\mathbf{hh}}_{r,k}^* + \textstyle\sum\limits_{s=1}^S \sum\limits_{i=1}^K \eta_{is} \big{[} (b_{d,is}^2+N \alpha_{1,is}+N \alpha_{3,is})\notag\\
&\overline{\mathbf{hh}}_{r,k}^T \overline{\mathbf{GG}_s}^T \overline{\mathbf{GG}}_s^* \overline{\mathbf{hh}}_{r,k}^* + (b_{d,ks}^2+N \alpha_{1,ks}+N \alpha_{3,ks})\overline{\mathbf{hh}}_{r,i}^T \overline{\mathbf{GG}_s}^T \overline{\mathbf{GG}}_s^* \overline{\mathbf{hh}}_{r,i}^*  \big{]} \bigg{\}},
\end{align}
\vspace{-15mm}
\begin{align}
&\mathbf{t}=\textstyle\sum\limits_{k=1}^K \kappa_k \mu_k |r_k|^2 \bigg{\{} \sum\limits_{s=1}^S \sum\limits_{t=1}^S \sum\limits_{i=1}^K \sqrt{\eta_{is} \eta_{it}} \big{[} \overline{\mathbf{hh}}_{r,k}^T\overline{\mathbf{GG}}_s^T (\overline{\mathbf{GG}}_s \overline{\mathbf{hh}}_{r,i} \mathbf{x} + \overline{\mathbf{hh}}_{d,is})^* (\overline{\mathbf{GG}}_t \overline{\mathbf{hh}}_{r,i} \mathbf{x} + \overline{\mathbf{hh}}_{d,it})^T \notag\\
&\overline{\mathbf{hh}}_{d,kt}^* + \overline{\mathbf{hh}}_{r,k}^T \overline{\mathbf{GG}}_s^T \overline{\mathbf{GG}}_{is}^* \overline{\mathbf{GG}}_{it}^T  \overline{\mathbf{hh}}_{d,kt}^* + \overline{\mathbf{hh}}_{r,i}^T \overline{\mathbf{GG}}_t^T \overline{\mathbf{GG}}_{kt}^* \overline{\mathbf{GG}}_{ks}^T  \overline{\mathbf{hh}}_{d,is}^* + M \overline{\mathbf{hh}}_{r,k}^T \overline{\mathbf{GG}}_s^T (\overline{\mathbf{GG}}_s \overline{\mathbf{hh}}_{r,i} \notag\\
&\mathbf{x} + \overline{\mathbf{hh}}_{d,is})^*\widehat{\mathbf{hh}}_{r,it} \widehat{\mathbf{hh}}_{r,kt}^H + M \overline{\mathbf{hh}}_{r,k}^T \overline{\mathbf{GG}}_s^T \widehat{\mathbf{GG}}_{is}^* \overline{\mathbf{hh}}_{r,kt}^* \mathbf{x}^*+M \overline{\mathbf{hh}}_{r,i}^T \overline{\mathbf{GG}}_s^T \widehat{\mathbf{GG}}_{ks}^* \overline{\mathbf{hh}}_{r,it}^* \mathbf{x}^* \big{]}+2M\textstyle \sum\limits_{s=1}^S\notag\\
& \textstyle\sum\limits_{t=1}^S \sqrt{\eta_{ks} \eta_{kt}} (b_{d,kt}^2+N\alpha_{2,kt}+N\alpha_{3,kt})\overline{\mathbf{hh}}_{r,k}^T \overline{\mathbf{GG}_s}^T \overline{\mathbf{hh}}_{d,ks}^* + \sum\limits_{s=1}^S \sum\limits_{i=1}^K \eta_{is} \big{[} (b_{d,is}^2+N \alpha_{1,is}+N \alpha_{3,is})\overline{\mathbf{hh}}_{r,k}^T \notag\\
& \overline{\mathbf{GG}_s}^T  \overline{\mathbf{hh}}_{d,ks}^* + (b_{d,ks}^2+N \alpha_{1,ks}+N \alpha_{3,ks})\overline{\mathbf{hh}}_{r,i}^T \overline{\mathbf{GG}_s}^T  \overline{\mathbf{hh}}_{d,is}^* \big{]} -\frac{1}{2}\textstyle\sum\limits_{s=1}^S \eta_{ks} (b_{d,ks}^2+N \alpha_{1,ks}+N \alpha_{3,ks})\notag\\
&\overline{\mathbf{hh}}_{r,k}^T \overline{\mathbf{GG}_s}^T (\overline{\mathbf{GG}}_s \overline{\mathbf{hh}}_{r,k} \mathbf{x}+\overline{\mathbf{hh}}_{d,ks})^*  \bigg{\}} - \textstyle\sum\limits_{k=1}^K \kappa_k \mu_k r_k \sum\limits_{s=1}^S \sqrt{\eta_{ks}} \overline{\mathbf{hh}}_{r,k}^T \overline{\mathbf{GG}_s}^T (\overline{\mathbf{GG}}_s \overline{\mathbf{hh}}_{r,k} \mathbf{x}+\overline{\mathbf{hh}}_{d,ks})^*,
\end{align}
\vspace{-15mm}
\begin{align}
&\mathbf{Q}=\textstyle\sum\limits_{k=1}^K \kappa_k \mu_k |r_k|^2  \sum\limits_{s=1}^S \sum\limits_{t=1}^S \sum\limits_{i=1}^K \sqrt{\eta_{is} \eta_{it}} \overline{\mathbf{hh}}_{r,i}^T\overline{\mathbf{GG}}_t^T (\overline{\mathbf{GG}}_t \overline{\mathbf{hh}}_{r,k} \mathbf{u} + \overline{\mathbf{hh}}_{d,kt})^* (\overline{\mathbf{GG}}_s \overline{\mathbf{hh}}_{r,k}  \mathbf{u} + \overline{\mathbf{hh}}_{d,ks})^T\notag\\
& \overline{\mathbf{GG}}_s^* \overline{\mathbf{hh}}_{r,i}^*
\end{align}
and\vspace{-5mm}
\begin{align}
&\mathbf{a}=\textstyle\sum\limits_{k=1}^K \kappa_k \mu_k |r_k|^2 \bigg{\{} \sum\limits_{s=1}^S \sum\limits_{t=1}^S \sum\limits_{i=1}^K \sqrt{\eta_{is} \eta_{it}} \big{[} \overline{\mathbf{hh}}_{r,i}^T\overline{\mathbf{GG}}_t^T (\overline{\mathbf{GG}}_t \overline{\mathbf{hh}}_{r,k}\mathbf{u} + \overline{\mathbf{hh}}_{d,kt})^* (\overline{\mathbf{GG}}_s \overline{\mathbf{hh}}_{r,k} \mathbf{u} + \overline{\mathbf{hh}}_{d,ks})^T \notag\\
&\overline{\mathbf{hh}}_{d,is}^*+ M \overline{\mathbf{hh}}_{r,i}^T \overline{\mathbf{GG}}_s^T (\overline{\mathbf{GG}}_s \overline{\mathbf{hh}}_{r,k} \mathbf{u} + \overline{\mathbf{hh}}_{d,ks})^* \widehat{\mathbf{hh}}_{r,kt} \widehat{\mathbf{hh}}_{r,it}^H + M \overline{\mathbf{hh}}_{r,kt}^T \widehat{\mathbf{GG}}_{is}^T (\overline{\mathbf{GG}}_s \overline{\mathbf{hh}}_{r,k} \mathbf{u}+\overline{\mathbf{hh}}_{d,ks})^*\notag\\
&+ M \overline{\mathbf{hh}}_{r,it}^T \widehat{\mathbf{GG}}_{ks}^T (\overline{\mathbf{GG}}_s \overline{\mathbf{hh}}_{r,i} \mathbf{u}+\overline{\mathbf{hh}}_{d,is})^*  \big{]} -\frac{1}{2}\textstyle\sum\limits_{s=1}^S \eta_{ks} (b_{d,ks}^2+N \alpha_{1,ks}+N \alpha_{3,ks})\overline{\mathbf{hh}}_{r,k}^T \overline{\mathbf{GG}_s}^T(\overline{\mathbf{GG}}_s\notag\\
& \overline{\mathbf{hh}}_{r,k} \mathbf{u}+\overline{\mathbf{hh}}_{d,ks})^*  \bigg{\}} - \textstyle\sum\limits_{k=1}^K \kappa_k \mu_k r_k \sum\limits_{s=1}^S \sqrt{\eta_{ks}} \overline{\mathbf{hh}}_{r,k}^T \overline{\mathbf{GG}_s}^T (\overline{\mathbf{GG}}_s \overline{\mathbf{hh}}_{r,k} \mathbf{u}+\overline{\mathbf{hh}}_{d,ks})^*.
\end{align}

\end{document}